\documentclass{elsart}

\usepackage{natbib}

\usepackage{amssymb}

\begin{document}

\begin{frontmatter}

\title{On the verge of {\textit {\textbf Umdeutung} in Minnesota: Van Vleck and the correspondence principle. Part Two}\thanksref{mpiwg}}
\thanks[mpiwg]{This paper was written as part of a joint project in the history of quantum physics of the {\it Max Planck Institut f\"{u}r Wissenschaftsgeschichte} and the {\it Fritz-Haber-Institut} in Berlin. The authors gratefully acknowledge support from the Max Planck Institute for History of Science. The research of Anthony Duncan is supported in part by the National Science Foundation under grant PHY-0554660.}

\author[duncan]{Anthony Duncan},
\author[janssen]{Michel Janssen\corauthref{cor}}
\corauth[cor]{Corresponding author. Address: Tate Laboratory of Physics, 116 Church St.\ NE, Minneapolis, MN 55455, USA, Email: janss011@tc.umn.edu}
\address[duncan]{Department of Physics and Astronomy, University of Pittsburgh}
\address[janssen]{Program in History of Science, Technology, and Medicine, University of Minnesota}

\begin{abstract}
 This is the second installment of a two-part paper on developments in quantum dispersion theory leading up to Heisenberg's {\em Umdeutung} paper. In telling this story, we have taken a paper by John H. \citet{Van Vleck 1924b, Van Vleck 1924c} as our main guide. In this second part we present the detailed derivations on which our narrative in the first part rests. The central result that we shall derive is the Kramers dispersion formula, which played a key role in the thinking that led to Heisenberg's {\em Umdeutung} paper. Closely following Van Vleck's pre-{\em Umdeutung} approach, we derive classical and construct quantum formulae for the dispersion, emission, and absorption of radiation both for the special case of a charged harmonic oscillator (sec.\ 5) and for arbitrary non-degenerate multiply-periodic systems (sec.\ 6). In sec.\ 7, we rederive the same results using modern quantum mechanics. In sec.\ 8 we bring together the main conclusions of our study.    
\end{abstract}


\end{frontmatter}

\setcounter{section}{4}
\setcounter{equation}{14}
\setcounter{footnote}{170}

\section{Van Vleck and the application of the correspondence principle to the interaction
of matter and radiation}

 In the two-part paper that forms the focal point of our study, \citet{Van Vleck 1924b,Van Vleck 1924c} explored in a systematic
  and   physically cogent fashion the implications of the correspondence principle for several aspects
  of the  interaction of 
matter and radiation. The paper is signed June 19, 1924 and appeared in the October 1924 issue of {\it The Physical Review}. In this paper, Van Vleck gives a detailed derivation of the correspondence
  principle for absorption, which he had introduced in a short note in the {\it Journal of the Optical Society in America}, signed April 7, 1924 \citep{Van Vleck 1924a}. In addition, he thoroughly 
examined the issues involved in connecting
  Einstein's $A$ and $B$ coefficients to features of classical electron orbits. Finally, as we mentioned in sec.\  3.4 in Part One of our paper,  he showed that, in the limit of high quantum numbers, Kramers' quantum formula for polarization merges with the  classical formula for polarization in arbitrary non-degenerate multiply-periodic systems.  

In part I of his paper,  reproduced in \citep{Van der Waerden}, \citet{Van Vleck 1924b} discusses the transition from quantum-theoretical expressions
   for emission, absorption, and dispersion to corresponding classical expressions that one 
   expects to hold in the limit of high quantum numbers. It is only in part II, not included in \citep{Van der Waerden}, that \citet{Van Vleck 1924c} derives the classical expressions for absorption and dispersion of radiation by a general non-degenerate 
   multiply-periodic system, using standard
   methods of canonical perturbation theory in action-angle variables. Van Vleck could assume his audience to be thoroughly familiar with these techniques. This is no longer true today. For the sake of clarity of exposition,
   we therefore invert the order of Van Vleck's own presentation. 
   
   In sec.\ 5.1, we present  the basic elements of
   the canonical formalism in action-angle variables and use it  to rederive the classical formula (6) in sec.\ 3.1
for the dipole moment of a charged one-dimensional simple harmonic oscillator. Though much more complicated than the derivation in sec.\ 3.1, this new derivation has two distinct advantages.  First, it suggests a way of translating the classical formula into a quantum formula with the help of Bohr's correspondence principle and Einstein's $A$ and $B$ coefficients. Secondly, both the derivation of the classical formula and its translation into a quantum formula can easily be generalized to arbitrary non-degenerate multiply-periodic systems. 
   
   In sec.\ 5.2, we translate the classical formula for the dipole moment of a simple harmonic oscillator into a quantum formula. In sec.\ 5.3, we  similarly convert classical formulae for emission and absorption by a simple harmonic oscillator to the corresponding quantum formulae. Both the mathematical manipulations and the physical interpretation are particularly transparent in the case of a simple harmonic oscillator, and Van Vleck himself frequently used this example for illustrative purposes. The generalization of the various results to arbitrary non-degenerate multiply-periodic systems, which is a primary focus of Van Vleck's paper, will be deferred to sec.\ 6. In sec. 7, we present (or outline) modern derivations of various results in secs.\ 5 and 6.

\subsection{Deriving the classical formula for the dipole moment of a simple harmonic oscillator using canonical perturbation theory}
\label{classical part}

In this subsection we rederive formula (6) in sec.\ 3.4
for the dipole moment of a charged one-dimensional simple harmonic oscillator, using canonical perturbation theory in action-angle variables. Like Kramers, Van Vleck was a master of these techniques in classical mechanics. As Van Vleck recalled fifty years after the fact:
\begin{quotation}
In 1924 I was an assistant professor at the University of Minnesota. On an American trip, Ehrenfest gave a lecture there \ldots [He] said he would like to hear a colloquium by a member of the staff. I was selected to give a talk on my ``Correspondence Principle for Absorption" \ldots I remember Ehrenfest being surprised at my being so young a man. The lengthy formulas for perturbed orbits in my publication on the three-body of the helium atom [Van Vleck, 1922b] had given him the image of a venerable astronomer making calculations in celestial mechanics \citep[p.\ 9]{Van Vleck 1974}.\footnote{Van Vleck failed to conform to Ehrenfest's image of a young physicist in another respect. In an interview in 1973, ``Van Vleck recalled, ``I shocked Ehrenfest \ldots when I told him I liked popular music." Ehrenfest, he said, ``thought that was completely irreconcilable with my having written any respectable papers."" \citep[p.\ 54]{Fellows}}
\end{quotation}

We begin by reviewing some of the mathematical tools we need.\footnote{This material is covered in standard graduate textbooks on classical mechanics, such as
\citep{Goldstein}, heavily influenced by \citep{Born 1925} \citep[pp.\ 429, 493, 540]{Goldstein}. We recommend \citep{Matzner}.}
Consider a classical Hamiltonian system with phase space coordinates
$(q_i,p_i)$, $i=(1,2,\ldots N)$ and Hamiltonian $H(q_i,p_i)$, which does not explicitly depend on time. Hamilton's equations are
\begin{equation}
\label{eq:2.1}
    \dot{q}_i = \frac{\partial H}{\partial p_i},\;\;\;
    \dot{p}_{i} = -\frac{\partial H}{\partial q_{i}}.
\end{equation}
Consider a contact transformation
$ (q_{i},  p_i) \rightarrow (q^\prime_i, p_{i}^{\prime})$
preserving the
form of  Hamilton's equations, in the sense that there exists a new Hamiltonian
$H^{\prime}$ such that 
\begin{eqnarray}
\label{eq:2.2}
    \dot{q}^\prime_i   = \frac{\partial H^{\prime}}{\partial p^{\prime}_i},\;\;
    \dot{p}^{\prime}_i = -\frac{\partial H^{\prime}}{\partial q^\prime_i}.
\end{eqnarray}
Since Hamilton's equations (\ref{eq:2.1}) and (\ref{eq:2.2}) must hold simultaneously,
the variational principles
\begin{equation}
\label{eq:2.3}
  \delta\int_{t_1}^{t_2}  \left( \sum_i p_{i}\dot{q}_i -H(q_i,p_i) \right)dt = 0, \;\;
  \delta\int_{t_1}^{t_2} \left( \sum_i p^{\prime}_ i \dot{q}^\prime_i  
-H^{\prime}(p^{\prime}_{i},q^\prime_i) \right)dt = 0
\end{equation}
for arbitrary times $t_1$ and $t_2$ must also hold simultaneously. This implies that the difference between the two integrands in eq.\ (\ref{eq:2.3}) must be a total time derivative
\begin{equation}
\label{eq:2.4}
  \left( \sum_i p_{i}\dot{q}_i-H(q_i,p_i) - \sum_i p^{\prime}_{i}\dot{q}^\prime_i  
+H^{\prime}(p^{\prime}_{i},q^\prime_i) \right) dt = dF,
\end{equation}
which will not contribute to the variation of the action.
The apparent dependence of $F$ on the $4N+1$ variables $(q_i,p_i,q^\prime_i,p^{\prime}_i,t)$
can be reduced to $2N+1$ variables via the equations for the contact transformation $ (q_{i},  p_i) \rightarrow (q^\prime_i, p_{i}^{\prime})$. If we choose to
write $F$ as a function of the initial and final coordinates, $F=F(q_i,q^\prime_i,t)$, then the
partial derivatives of $F$ can be read off directly from eq.\ (\ref{eq:2.4}):
\begin{equation}
\label{eq:2.5}
   \frac{\partial F}{\partial t} = H^{\prime}-H,  \;\;\;\;
   \frac{\partial F}{\partial q_i} = p_i, \;\;\;\;
   \frac{\partial F}{\partial q^\prime_i} = -p^{\prime}_{i}. \nonumber
\end{equation}
By solving  (at least in principle!) the second of these three equations
for $q^\prime_i$ as a function of $(q_i,p_i)$, and then substituting
the result in the third to obtain $p^{\prime}_i$, we see that the function $F$ encodes
the full information of the transformation $ (q_{i},  p_i) \rightarrow (q^\prime_i, p_{i}^{\prime})$. This function is called the {\it generating function} of the transformation. Given $F$ the form of the new Hamiltonian $H^{\prime}$ can be
obtained (again, in principle!) from the first of eqs.\ (\ref{eq:2.5}).

  A special case of great interest occurs when the generating function $F$ can be
chosen so that the resulting Hamiltonian is  independent of the new coordinates
$q^\prime_i$ (which are then called {\it ignorable}). Hamilton's equations 
then immediately imply that the associated momenta $p^{\prime}_{i}$ are time-independent,
and that the new coordinates $q^{\prime}_{i}$ are linear in time. In
this circumstance the new momenta are usually called {\it action variables}---the notation
$J_{i}$ is conventional for these---while the new coordinates are dubbed {\it angle variables},
with the conventional notation $w_{i}$.

To illustrate the above with a concrete example, which we shall be using throughout this section, 
consider a one-dimensional simple harmonic oscillator with Hamiltonian:\footnote{A short
digression on the (almost inevitable) notational confusions lurking in this subject
is in order. We shall continue to use the conventional notation $\omega$ to denote
angular frequencies, with the ordinary frequency (reciprocal period) denoted by the
Greek letter $\nu$. Unfortunately, Van Vleck uses $\omega$ to denote ordinary
frequency! Moreover, there is the embarrassing similarity of the angle variables $w_i$
to the frequencies $\omega_i$. Also, there is the need to distinguish between the
frequencies of the isolated mechanical system ($\omega_0=2\pi\nu_{0}$ for the simple harmonic oscillator) and the
frequency of an applied electromagnetic wave, which we shall denote as $\omega = 2\pi\nu$
throughout.}
\begin{equation} 
\label{eq:2.6}
   H= \frac{p^{2}}{2m} +\frac{1}{2}m\omega_{0}^{2}q^{2}.
\end{equation}
Consider the transformation
induced by
\begin{equation}
\label{eq:2.7}
  F = \frac{1}{2}m\omega_{0}q^{2}\cot{q^{\prime}}.
\end{equation}
This function does not explicitly depend on time, so $H^{\prime} = H$ (see eq.\ (\ref{eq:2.5})). Eq.\ (\ref{eq:2.5}) also tells us that
\begin{equation}
  p = \frac{\partial F}{\partial q} = m\omega_{0}q\cot{q^{\prime}}, \;\;\;\;
  p^{\prime} = -\frac{\partial F}{\partial q^{\prime}} = \frac{1}{2}m\omega_{0}q^{2}\csc^{2}{q^{\prime}}.
\end{equation}
From the latter equation it follows that $ q^{2}=({2p^{\prime}}/{m\omega_{0}})\sin^{2}{q^{\prime}}$ or that
\begin{equation}
\label{eq:2.8}
q=\sqrt{\frac{2p^{\prime}}{m\omega_{0}}}
\sin{q^{\prime}}.
\end{equation}
Inserting this expression for $q$ into the expression for $p$, we find 
\begin{equation}
\label{eq:2.9}
 p = \sqrt{2m\omega_{0}p^{\prime}}\cos{q^{\prime}}.
\end{equation} 
Substituting eqs.\ (\ref{eq:2.8})--(\ref{eq:2.9}) for $q$ and $p$ into eq.\ (\ref{eq:2.6}) we find
\begin{equation}
\label{eq:2.9a}
H = \omega_{0}p^{\prime}.
\end{equation}
Since $H^{\prime} = H$, this means that the new coordinate variable $q^{\prime}$ is {\it ignorable}, as desired. Hamilton's equations for $(q^{\prime}, p^{\prime})$ are:
\begin{equation}
\label{eq:2.10}
  \dot{q}^{\prime} = \frac{\partial H}{\partial p^{\prime}} = \omega_{0}, \;\;\;
  \dot{p}^{\prime} = - \frac{\partial H}{\partial q^\prime}=0,
\end{equation}
from which it follows that $q^{\prime}=\omega_{0}t+\epsilon$ and that $p^{\prime}=H/\omega_{0}$ is time-independent. 
Instead of the canonically conjugate variables ($p^{\prime},q^{\prime}$) it is customary to employ 
rescaled action/angle variables
\begin{equation}
\label{eq:2.11}
  J \equiv 2\pi p^{\prime},\;\;\; w \equiv \frac{1}{2\pi} q^{\prime}. 
 \end{equation}
 Hamilton's equations for $(J, w)$ are:
 \begin{equation}
 \dot{w} = \frac{\partial H}{\partial J} = \nu_{0}, \;\;\;
  \dot{J} = - \frac{\partial H}{\partial w}=0.
\end{equation}
It follows that $J=H/\nu_{0}$ and $w=\nu_{0}t+\epsilon$ (appropriately redefining the arbitrary
phase $\epsilon$) for our one-dimensional oscillator. 

The connection to the terminology  {\it action variable}
is easily seen in this example. In this simple case, the action  is defined as the area enclosed by a single orbit of the periodic system in the two-dimensional phase space spanned by the coordinates $(p, q)$:
\begin{equation}
\label{eq:2.12}
 J = \oint p dq.
 \end{equation}
Inserting eqs.\ (\ref{eq:2.8}) and (\ref{eq:2.9}) into the integrand, we find
\begin{equation}
\label{eq:2.13}
\oint \left( \sqrt{2 m \omega_0 p^{\prime}}\cos{q^{\prime}} \right) d\left( \sqrt{\frac{2p^{\prime}}{m\omega_{0}}}
\sin{q^{\prime}} \right) = \int_{0}^{2\pi} 2p^{\prime}\cos^{2}{q^{\prime}}dq^{\prime} = 2\pi p^{\prime},
\end{equation}
which is just the expression for $J$ in eq.\ (\ref{eq:2.11}).

The result (\ref{eq:2.8}) represents, of course, the solution of the equation of 
motion of the oscillator
\begin{equation}
\label{shosolution}
q(t)= D\cos{2\pi\nu_{0}t} = D\cos{2\pi w},
 \end{equation}
where we 
have chosen the phase shift $\epsilon$ to start the oscillator at maximum displacement
at $t=0$, and where the amplitude is a function of the action variable
\begin{equation}
\label{eq:2.13a}
 D=\sqrt{\frac{J}{m\pi \omega_{0}}}.
 \end{equation}


We now turn to our basic model for dispersion, i.e., a charged one-dimensional simple harmonic oscillator subjected to the periodically varying electric field of an electromagnetic wave.
Earlier, we used elementary techniques of classical mechanics to analyze this system (see eqs.\ (2)--(6) in sec.\ 3.1).
Although such methods are physically
  transparent, they depend on an explicit treatment of the equations of motion of a specific and
  completely specified Hamiltonian. The same results can be obtained by the methods
  of canonical perturbation theory, where general formulas can be obtained for the perturbation
  in the coordinate(s) of the system {\it completely independently of the specific nature of the
  dynamics}.   
  As Van Vleck put it:
  \begin{quotation}
  If we were to study the perturbations in the motion produced by the incident wave purely with the aid of [Newton's second law] it would be impossible to make further progress without specializing the form of the potential function [such as, e.g., $\frac{1}{2} m \omega_0^2 q^2$ in eq.\ (\ref{eq:2.6})] \ldots However, it is quite a different story when we seek to compute the perturbations \dots in the ``angle variables" $w_1$, $w_2$, $w_3$ and their conjugate momenta $J_1$, $J_2$, $J_3$ \ldots In fact by using them rather than $x,y,z$, which is the essential feature of the present calculation, the periodic properties of the system come to light even without knowing the form of [the potential]  \citep[p.\ 350]{Van Vleck 1924c}.
  \end{quotation}
Using  canonical perturbation theory in action-angle variables, we rederive eq.\ (6) 
of sec.\ 3.1 for the polarization of a one-dimensional charged simple harmonic oscillator. In sec.\ 6,  we turn to the  general case of an arbitrary non-degenerate multiply-periodic system.

 The Hamiltonian is now the sum of the Hamiltonian $H_0$ given by eq.\ (\ref{eq:2.6}) and a perturbative term $H_{\rm{int}}$ describing the interaction between the harmonic oscillator and the electromagnetic wave:\footnote{As before, we assume that the electric field is in the direction of motion of the oscillator (cf.\ sec.\ 3.1).
It follows from eq.\ (\ref{eq:2.19}) that the force $F=-\partial V/\partial x$ of the electric field on the charge is $- e E \cos{\omega t}$, in accordance with eq.\ (3) in sec.\ 3.1
(recall that we use $e$ to denote the absolute value of the electron charge). }
\begin{equation}
\label{eq:2.19}
H=H_0 + H_{\rm{int}}=\frac{p^2}{2m} + \frac{1}{2}m\omega_0^2 x^2 + e E x \cos{\omega t}.
\end{equation}
The subscript `0' in $\nu_0$ or $\omega_0$ refers to the characteristic frequency of the unperturbed oscillator. Without subscript $\nu$ and $\omega$ refer to the frequency of the external electric field.

Absent a perturbing field ($E=0,H=H_0$),
  we can write $x(t)$ in terms of the action-angle variables $J$ and $w=\nu_0 t$:
  \begin{equation}
    \label{eq:2.20}
  x(t) =  \sum_{\tau=\pm1}A_{\tau}(J)e^{2\pi i \tau w},
   \end{equation}
where $A_\tau$ has to satisfy the conjugacy relation $A_{\tau}=A_{-\tau}^{*}$ to ensure that $x(t)$ in eq.\ (\ref{eq:2.20})
  is real ($x(t)=x^*(t)$). Note that we have changed notation somewhat compared to eq.\ (\ref{shosolution}). We returned to Cartesian coordinate notation ($x$ instead
of $q$), and the amplitude  has been redefined:\footnote{Inserting $A_\tau = |A_\tau| e^{i \varphi}$ into eq.\ (\ref{eq:2.20}), we find $x(t) = \left( |A_\tau| + |A_{-\tau}| \right) \cos{(2 \pi w + \varphi)}$. Since $A_{\tau}=A_{-\tau}^{*}$, $|A_\tau|^2 = A_\tau A^*_\tau$ is equal to $|A_{-\tau}|^2 = A_{-\tau} A^*_{-\tau}$. The phase angle $\varphi$ is immaterial. \label{amp}} 
 \begin{equation} 
  D = 2|A_{\tau}|. 
  \label{eq:2.20a}
\end{equation}
The action-angle variables
 $J= H_0/\nu_0$ and $w=\nu_0 t$ satisfy Hamilton's equations (cf.\ eqs.\ (\ref{eq:2.10})--(\ref{eq:2.11})):
  \begin{equation}
  \label{eq:2.21}
  0 =  -\dot{J} = \frac{\partial H_0}{\partial w},  \;\;\;\;\;\;
    \frac{\partial H_0}{\partial J} =\dot{w}=\nu_0.
  \end{equation}
  It is a special feature of the simple harmonic oscillator that the frequency $\nu_0$ is independent of
  the amplitude of motion (and thereby of the action). The generating function for the 
  contact transformation from $(x,p)$ to $(w,J)$ is time-independent (cf.\ eq.\ (\ref{eq:2.7})),
  so eq.\ (\ref{eq:2.5}) implies
  that the old and new Hamiltonians coincide in value (i.e., one simply reexpresses the
  original Hamiltonian in the new variables). Even with the perturbation turned on
  {\it we shall continue to use the same contact transformation}, computing the 
  perturbations $(\Delta w,\Delta J)$ induced by the applied field in the action-angle variables $(w,J)$ as 
  an expansion in $E$. These are {\it not}  action-angle variables for the full Hamiltonian $H_0 + H_{\rm int}$, only for the unperturbed Hamiltonian $H_0$ (cf.\ Van Vleck 1926a, pp.\ 200--201).
  
  Eventually, we are interested in the displacement  $\Delta x$ in
  the particle coordinate (to first order in $E$) induced by the applied field. To first order, $\Delta x$ is given by
  \begin{equation}
  \label{eq:2.22}
     \Delta x = \frac{\partial x}{\partial J}\Delta J + \frac{\partial x}{\partial w}\Delta w.
     \end{equation} 
Using eq.\ (\ref{eq:2.20}) to evaluate $\partial x/\partial J$ and $\partial x/\partial w$, we can rewrite this as:
 \begin{eqnarray}
  \label{eq:2.28}
   \Delta x = \sum_{\tau} \left(\frac{\partial A_{\tau}}{\partial J}\Delta J+2\pi i \tau A_{\tau} \Delta w\right)e^{2\pi i \tau w}. 
   \end{eqnarray}
Assuming the external field to be switched on at time zero, the first-order shifts $\Delta w$ and $\Delta J$ are given by:
 \begin{equation}
 \Delta J = \int_{0}^{t} \Delta \dot{J} dt, \;\;\; \Delta w = \int_{0}^{t} \Delta \dot{w} dt.
 \label{eq:2.22a}
 \end{equation}
 where the integrands $\Delta \dot{J}$ and $\Delta \dot{w}$ are determined by Hamilton's equations.

The perturbation in eq.\ (\ref{eq:2.19})  will induce
  a time-dependence in the action variable, as  Hamilton's equation for the action variable in the
  presence of the perturbing field now reads
   \begin{equation}
  \label{eq:2.23}
  \dot{J}= - \frac{\partial H_0}{\partial w} - eE\frac{\partial x}{\partial w}\cos{2\pi\nu t}=
  - eE\frac{\partial x}{\partial w}\cos{2\pi\nu t}.
  \end{equation}
Note that we still have $\partial H_0/\partial w=0$, so $\Delta \dot{J} = \dot{J}$. At this point it is convenient to go over to complex
  exponentials and replace $\cos{2\pi\nu t}$ by $\frac{1}{2}(e^{2\pi i\nu t}+e^{-2\pi i\nu t})$.
  Inserting eq.\ (\ref{eq:2.20}) into eq.\ (\ref{eq:2.23}), we find
  \begin{equation}
  \label{eq:2.24}
\Delta \dot{J}=-\pi ieE\sum_{\tau=\pm1}\tau A_{\tau}\left(e^{2\pi i(\tau w+\nu t)}+e^{2\pi i(\tau w-\nu t)}\right).
   \end{equation}
   To obtain the polarization, which is a linear effect in the applied field $E$, we
   only need $\Delta J$ and $\Delta w$ to first order in $E$. This means that the angle variables
   $w$ in the exponents in eq.\ (\ref{eq:2.24}) can be taken to zeroth order, i.e., $w=\nu_{0}t$.
   Integrating $\Delta \dot{J}$ we find:
   \begin{equation}
   \label{eq:2.25}
    \Delta J =\int_{0}^{t} \Delta \dot{J}dt= \frac{eE}{2}\sum_{\tau=\pm1}\tau A_{\tau} \left\{ \frac{1-e^{2\pi i(\tau\nu_{0}t+\nu t)}}{\tau\nu_{0}+\nu}+\frac{1-e^{2\pi i(\tau\nu_{0}t-\nu t)}}{\tau\nu_{0}-\nu} \right\}.
    \end{equation}
  Next, we need to compute the first order shift $\Delta w$ in the angle variable $w$. Hamilton's
  equation for the angle variable $w$ in the presence of the perturbation is:\footnote{\label{nu_0(J)} It is a special feature of the simple harmonic oscillator that the characteristic frequency $\nu_0$ is independent of the amplitude and thus of the action variable $J$ (see eq.\ (\ref{eq:2.13a})). In general, $\nu_0$ will be a function of $J$. The first term on the right-hand side of eq.\ (\ref{eq:2.26}) would then become $\partial H_{0} / \partial J = \nu_{0}(J) = \nu_0 + (\partial \nu_0 / \partial J) \Delta J$.}
  \begin{eqnarray}
  \label{eq:2.26}
   \dot{w} &=& \frac{\partial H_{0}}{\partial J}+eE\frac{\partial x}{\partial J}\cos{2\pi\nu t} \nonumber \\
     &=& \nu_{0}+\frac{eE}{2}\sum_{\tau=\pm1}\frac{\partial A_{\tau}}{\partial J} \left(e^{2\pi i(\tau w+\nu t)}
     +e^{2\pi i(\tau w -\nu t)}\right).  
    \end{eqnarray}
   Once again, $w$ may be replaced by $\nu_{0}t$ in the exponentials in eq.\ (\ref{eq:2.26}).
   Integrating the second term in eq.\ (\ref{eq:2.26}), which gives the shift $\Delta \dot{w}$ due to $H_{\rm int}$,
we find:
   \begin{equation}
   \label{eq:2.27}
    \Delta w = \int_{0}^{t} \Delta\dot{w}dt=  \frac{ieE}{4\pi}\sum_{\tau=\pm1}\frac{\partial A_{\tau}}{\partial J}
   \left\{ \frac{1-e^{2\pi i(\tau\nu_{0}t+\nu t)}}{\tau\nu_{0}+\nu}+\frac{1-e^{2\pi i(\tau\nu_{0}t-\nu t)}}{\tau\nu_{0}-\nu}
   \right\}.
    \end{equation}
 Substituting expressions (\ref{eq:2.25}) and (\ref{eq:2.27}) for $\Delta J$ and $\Delta w$ into eq.\ (\ref{eq:2.28}), we find
   \begin{eqnarray}
   \label{eq:2.29}
  \Delta x  & = & \frac{eE}{2} \sum_{\tau^{\prime}=\pm1} \sum_{\tau=\pm1}
  \left\{ 
 \frac{\partial A_{\tau^{\prime}}}{\partial J}\tau A_{\tau}
  - \tau^{\prime} A_{\tau^{\prime}}  \frac{\partial A_{\tau}}{\partial J}
  \right\} 
  \frac{1-e^{2\pi i(\tau\nu_{0}t-\nu t)}}{\tau\nu_{0}-\nu}e^{2\pi i\tau^{\prime}\nu_{0}t}  \\ & & 
 \;\;\;\;\;\;\; \;\;\;\;\;\;\;  \;\;\;\;\;\;\;  \;\;\;\;\;\;\;  + \;\; (\nu \rightarrow -\nu), \nonumber
    \end{eqnarray}    
where ``$(\nu \rightarrow -\nu)$" here and below is shorthand for: ``the same term with $\nu$ replaced by $- \nu$ everywhere." The coherent contribution to the polarization comes from the terms in eq.\ (\ref{eq:2.29})
  with the same time-dependence as the applied field, i.e., from terms in which the time-dependence is given by the factor $e^{\pm2\pi i \nu t}$. In the terminology of \citet[p.\ 361]{Van Vleck 1924c}: ``the part of the displacement which is resonant to the impressed wave." These are the terms in which the summation indices, which in the case of the simple harmonic oscillator only take on the values $\pm 1$, have opposite values, i.e.,  $\tau = -\tau^{\prime}$. The contribution of such terms to the first-order displacement is
  \begin{eqnarray}
    \Delta x_{\rm coh} &=&\frac{eE}{2}\sum_{\tau=\pm1}
    \left\{ \left(
    \frac{\partial A_{-\tau}}{\partial J}\tau A_{\tau}
    +
    \tau A_{-\tau} \frac{\partial A_{\tau}}{\partial J} \right)
    \frac{-e^{-2\pi i\nu t}}{\tau\nu_{0} - \nu}+(\nu\rightarrow -\nu) \right\} \nonumber \\
    \label{eq:2.30}
      &=& \frac{eE}{2}\sum_{\tau=\pm1}\tau \frac{\partial |A_{\tau}|^{2}}{\partial J} \left\{\frac{e^{-2\pi i\nu t}}{\nu - \tau\nu_{0}} - \frac{e^{2\pi i \nu t}}{ \nu + \tau\nu_{0}}\right\}.
      \end{eqnarray}
  The imaginary part of this expression  is a sum over the product of odd and
  even functions of the index $\tau$,
  \begin{equation}
  \label{eq:2:30a}
   -\frac{eE}{2}  \sum_{\tau=\pm 1}\tau\frac{\partial |A_{\tau}|^{2}}{\partial J} \left(\frac{1}{\nu - \tau\nu_{0}}+\frac{1}{\nu + \tau\nu_{0}}\right) \sin{2\pi\nu t},
     \end{equation}
     and therefore vanishes, leaving only the real part:
 \begin{eqnarray}
   \Delta x_{\rm coh} & = &  \frac{eE}{2} \sum_{\tau}\tau  \frac{\partial |A_{\tau}|^{2}}{\partial J}
   \;  \left(\frac{1}{\nu - \tau\nu_{0}} - \frac{1}{\nu + \tau\nu_{0}}\right) \cos{2\pi\nu t} \nonumber \\  & = &
   \frac{eE}{2} \sum_{\tau}\tau  \frac{\partial |A_{\tau}|^{2}}{\partial J}
    \left( \frac{2\tau\nu_{0}}{\nu^{2} - \tau^2\nu_{0}^{2}} \right) \cos{2\pi\nu t}.
   \label{eq:2.31a}
   \end{eqnarray}
  Since $|A_\tau|^2 = |A_{-\tau}|^2$ (see note \ref{amp}) and since $\tau$ only takes on the values $\pm1$ in the case of the simple harmonic oscillator, $\tau^2 =1$ and the two terms in the summation over $\tau$ are identical.  Although in this special case the derivative with respect to $J$ only acts on $|A_\tau|^2$, we are free to include the expression $2 \nu_0/(\nu^2 - \nu_0^2)$  within the scope of the derivative (recall that $\nu_0$ does not depend on $J$ in this case). Eq.\ (\ref{eq:2.31a}) then becomes     \begin{equation}
 \Delta x_{\rm coh}  = 2eE\frac{\partial}{\partial J}\left(\frac{\nu_{0}}{\nu^{2} - \nu_{0}^{2}}|A_{\tau}|^{2} \right) \cos{2\pi\nu t}.
 \label{eq:2.31}
   \end{equation} 

The resulting expression for the dipole moment, $p(t) = -e \Delta x_{\rm coh}$, of a one-dimensional charged simple harmonic oscillatoris a special case of the expressions for the dipole moment  of a general non-degenerate multiply-periodic system with the same charge given by Kramers and Van Vleck. 
 \citet[p.\ 310, eq.\ 2$^{*}$]{Kramers 1924b} denotes this quantity by $P$ and gives the following formula:
\begin{equation}
P = \frac{E}{2} \sum \frac{\partial}{\partial I} \left( \frac{C^2 \omega}{\omega^2 - \nu^2} \right) \cos{2 \pi \nu t}.
\label{Kramers}
\end{equation}
In the special case of a one-dimensional charged simple harmonic oscillator, $\omega$, $I$, and $C$ correspond to $\nu_0$, $J$, and $2|A_{\tau}|$ in our notation, respectively. There appears to be a factor $e^2$ missing in Kramers' formula. We shall derive the corresponding formula (41) in \citep[p.\ 361]{Van Vleck 1924c} in sec.\ 6.2.

Eq.\ (\ref{eq:2.31}) is equivalent to eq.\ (6)
the result of our much simpler derivation in sec.\ 3.1. Recalling that (cf.\ eqs.\ (\ref{shosolution})--(\ref{eq:2.13a}), eqs.\ (\ref{eq:2.20})--(\ref{eq:2.20a}) and note \ref{amp})
  \begin{equation}
  \label{eq:2.32}
  x(t)=2|A_{\tau}|\cos{2\pi\nu_{0}t}=\sqrt{\frac{J}{2 \pi^2 m\nu_{0}}}\cos{2\pi\nu_{0}t},
  \end{equation}
  we have $|A_{\tau}|^{2}=J/(8\pi^2 m \nu_{0})$, and eq.\ (\ref{eq:2.31}) reduces to
  \begin{equation}
    \Delta x_{\rm coh} = \frac{eE \cos{2\pi \nu t} }{ 4\pi^2m(\nu^{2} - \nu_{0}^{2}) }. 
  \end{equation}
 The dipole moment is thus given by:
\begin{equation}
p(t) = -e \Delta x_{\rm coh} = \frac{e^2E}{4 \pi^2 m (\nu_0^2 - \nu^2)} \cos{2 \pi \nu t},
\label{eq:2.32a}
\end{equation}
   in agreement with eq.\ (6) in sec.\ 3.1.
   
  The preceding discussion employs a version of canonical perturbation theory in which a single set of
     action-angle variables, chosen for the unperturbed Hamiltonian, is used throughout the calculation,
     even after the time-dependent perturbation is switched on. Accordingly, the new action variables 
     are no longer constant, and the new angle variables are no longer linear in time. The same classical polarization
     result is derived in a somewhat different manner by \citet{Born 1924} and by  \citet{Kramers and Heisenberg 1925}. Born performs a contact transformation in which the generating function $F$
     (cf. eq.\ (\ref{eq:2.4})) is chosen as a function of $(q_{i},p^{\prime}_{i})$, the old coordinates and the new momenta,
     which is then evaluated systematically order by order in the perturbation to maintain the constancy of the new
     action variables. In \citep{Kramers and Heisenberg 1925} the same procedure is followed, but as only the first order
     result is needed, it suffices to use the infinitesimal form of the contact transformation.\footnote{For a discussion of infinitesimal canonical transformations, see Ch.\ 11 of \citep{Matzner}.}
     


\subsection{Converting the classical formula for dispersion to a quantum formula in the special case of a simple harmonic oscillator}

Using Bohr's correspondence principle as our guide, we now `translate' the classical formula ({\ref{eq:2.31}) for displacement (and thence for polarization) into a quantum formula. Two main ingredients go into this particular application of the correspondence 
principle:  (1) a rule---commonly attributed to \citet{Born 1924}\footnote{See, e.g., 
\citep[p.\ 193]{Jammer 1966}, \citep[p. 148]{MacKinnon 1977}, \citep[pp.\ 178, 186, 188]{Cassidy}, or \citep[p.\ 1372]{Aitchison 2004}.}
but found and applied earlier by Kramers
and Van Vleck (see below)---for replacing derivatives with respect to the action variables in classical formulae by difference quotients involving neighboring quantum states;
(2) the $A$ and $B$ coefficients of Einstein's quantum theory of radiation. In general, the `translation' of a classical formula into a quantum formula involves a third step. The orbital frequencies need to replaced by transition frequencies. The case of a simple harmonic oscillator has the special features that the only relevant transitions are between adjacent states and that the transition frequency $\nu_{i \rightarrow f}$ coincides with the mechanical frequency $\nu_0$. Another special feature is that the correspondence between quantum and classical results for large quantum numbers continues to hold all the way down to the lowest quantum numbers, due to the extremely simple form of the energy spectrum, with uniformly spaced levels.

   Using the rule for replacing derivatives by difference quotients,
   the quantum  formula for polarization is obtained from (\ref{eq:2.31}) by the formal correspondence 
   replacement 
   \begin{equation}
   \label{eq:2.33}
  \left. \frac{\partial F}{\partial J}\right|_{J=rh}\rightarrow \frac{1}{h}(F(r+1)-F(r)),
   \end{equation}
    where $F(r)$ refers to any dynamical quantity associated with the quantum state specified by the
    integer quantum number $r$. In the correspondence limit where $r$ gets very large, the difference between the values $rh$ and $(r+1)h$ for the action variable $J$ become so small that the difference quotient to the right of the arrow in eq.\ (\ref{eq:2.33}) becomes equal the derivative on the left. With this prescription, the classical formula eq.\ (\ref{eq:2.31})
    turns into a quantum expression for the coherent part of the displacement of the particle in 
    quantum state $r$:
    \begin{equation}
    \label{eq:2.34}
    \Delta x_{\rm coh}^{r}= \frac{2eE}{h}\left(\frac{\nu_{0}|A^{r+1}|^{2}}{\nu^{2}-\nu_{0}^{2}}
    -\frac{\nu_{0}|A^{r}|^{2}}{\nu^{2}-\nu_{0}^{2}}\right) \cos{2\pi\nu t}.
    \end{equation}
The amplitudes $A^{r}$ correspond to the $A_{\tau}$ (with $\tau=\pm1$) in eq.\ (\ref{eq:2.31}),
 and are related to the amplitudes $D_{r}$ in eq.\ (\ref{eq:2.13a}) for an oscillator in state $r$ by $D_{r}=2|A^{r}|$
 (see eq.\ (\ref{eq:2.32})).
  As we saw in sec.\ 3.3, \citet{Ladenburg 1921} showed how these amplitudes can be connected to the Einstein $A$ coefficients  for spontaneous emission
  (not to be confused with the amplitudes $A^{r}$). 
  
  At this point we 
 briefly review Einstein's  quantum theory of radiation \citep{Einstein 1916a, Einstein 1916b, Einstein 1917a}, using the  notation of  \citep{Van Vleck 1924b}. 
  Imagine an ensemble of atoms---or indeed, any conceivable quantized mechanical
  system, such as one-dimensional quantized oscillators---in interaction and
  statistical equilibrium with an ambient electromagnetic field of spectral density
  $\rho(\nu)$. If we label the
  stationary states of the atoms by indices $r,s, \ldots$,  the number of atoms in
  state $r$ (of energy $E_r$) by $N_{r}$, and recall the Bohr frequency condition $\nu_{rs}=
  (E_r-E_s)/h$, Einstein's analysis gives an average rate of energy emission of
  light of frequency $\nu_{rs}$ for 
  an atom in state $r$ as
  \begin{equation}
  \label{eq:2.35}
    \frac{dE_{r\rightarrow s}}{dt} = h\nu_{rs} \left(A_{r\rightarrow s}+B_{r\rightarrow s}\rho(\nu_{rs})\right),
  \end{equation}
  and the rate of energy absorption of light of frequency $\nu_{rs}$ by an atom in state
  $s$ as
  \begin{equation}
  \label{eq:2.36}
   \frac{dE_{s\rightarrow r}}{dt} = h\nu_{rs}B_{s\rightarrow r}\rho(\nu_{rs}).
   \end{equation}
Einstein's analysis of the requirements for thermodynamic equilibrium and comparison with Planck's law of black-body radiation then yields
the critical relations
\begin{equation}
\label{eq:2.37}
  B_{r\rightarrow s}=B_{s\rightarrow r}=\frac{c^{3}}{8\pi h\nu_{rs}^{3}}A_{r\rightarrow s}.
\end{equation}   

 For a charged simple harmonic oscillator, the only allowed transitions amount to changes in
 the action by one unit of Planck's constant $h$, so there is only a single  Einstein coefficient for spontaneous 
 emission from the state $r+1$, namely $A_{r+1\rightarrow r}$.  The correspondence principle dictates
 that we associate the rate of spontaneous energy emission for {\it high} quantum numbers,
   \begin{equation}
  \label{eq:2.38}
    \frac{dE_{r+1\rightarrow r}}{dt} = h\nu_{0}A_{r+1\rightarrow r}
  \end{equation}
 (cf.\  eq.\ (\ref{eq:2.35}), in the absence of external radiation) with
 the classical result for the power emitted by an accelerated (in this case, oscillating) charge, given by
  the Larmor formula (Jackson, 1975; Feynman {\it et al.}, 1964, Vol.\ 1, Ch.\ 32):
 \begin{equation}
 \label{eq:2.39}
    P =\frac{2}{3}\frac{e^{2}}{c^{3}}\dot{v}^{2}.
   \end{equation}
For an oscillator in state $r$, with $x(t)=D_{r}\cos{\omega_{0}t}$, this becomes, for the instantaneous
power emission $P_{r}$ in state $r$
\begin{equation}
\label{eq:2.40}
  P_{r} =   \frac{2}{3}\frac{e^{2}}{c^{3}}\omega_{0}^{4}D_{r}^{2}\cos^{2}{\omega_{0}t},
  \end{equation}
 the time average of which, $\frac{1}{3}(e^{2}/c^{3})\omega_{0}^{4}D_{r}^{2}$,  then gives the desired connection between the amplitudes $D_{r} = 2 |A^{r}|$
 appearing in  eq.\ (\ref{eq:2.34}) and the Einstein coefficient $A_{r+1\rightarrow r}$:
 \begin{eqnarray}
   h\nu_{0}A_{r+1\rightarrow r} &=& \frac{4}{3}\frac{e^{2}}{c^{3}}\omega_{0}^{4}|A^{r+1}|^{2} \nonumber \\
   \label{eq:2.41}
   |A^{r+1}|^{2} &=& \frac{3hc^{3}}{64\pi^{4}e^{2}\nu_{0}^{3}}A_{r+1\rightarrow r}.
   \end{eqnarray}
\citet[p.\ 333]{Van Vleck 1924b} refers
 to this connection as the ``correspondence principle for emission." Multiplying the displacement $\Delta x$ in eq.\ (\ref{eq:2.34}) by the charge $-e$ to obtain the dipole moment
 per oscillator, and by $n_{\rm osc}$, the number density of oscillators, we obtain the following
 result for the polarization induced by the electric field $E$:
  \begin{equation}
  \label{eq:2.42}
   P_{r}=3\frac{n_{\rm osc}c^{3}}{32\pi^{4}}E \left( \frac{A_{r+1\rightarrow r}}{\nu_{0}^{2}(\nu_{0}^{2}-\nu^{2})}-\frac{A_{r\rightarrow r-1}}{\nu_{0}^{2}(\nu_{0}^{2}-\nu^{2})}\right) \cos{2\pi\nu t}.
   \end{equation}
   Of course, for the special case of the ground state of the oscillator, $r=0$, the second term in eq.\ (\ref{eq:2.42})
 cannot be present. Ladenburg's quantum formula for dispersion accordingly only had the equivalent of the first term in eq.\ (\ref{eq:2.42}) \citep{Ladenburg 1921}. The full equation corresponds to 
 eq.\ (5) in \citep{Kramers 1924a},  and to eq.\ (17) in \citep{Van Vleck 1924b}, except for a factor of 3, as we have not assumed
  random orientation of the oscillators \citep[footnote 25]{Van Vleck 1924b}.
  
 One may easily guess that the corresponding
   formula for a more general, multiply-periodic system will take the form of 
   \citep[eq (17)]{Van Vleck 1924b}, in analogy
   to (\ref{eq:2.42}) (cf. eqs.\ (6)--(7)):
   \begin{equation}
   \label{eq:2.43}
    P_{r} = 3\frac{n_{\rm osc}c^{3}}{32\pi^{4}}E \left( \sum_{s}\frac{A_{s\rightarrow r}}{\nu_{sr}^{2}(\nu_{sr}^{2}-\nu^{2})}-\sum_{t}\frac{A_{r\rightarrow t}}{\nu_{rt}^{2}(\nu_{rt}^{2}-\nu^{2})} \right) \cos{2\pi\nu t},
    \end{equation}
    where the sum over $s$ (resp. $t$) corresponds to states higher (resp. lower) than the
    state $r$, and where $\nu_{ij}$ is Van Vleck's notation for the transition frequency $\nu_{i \rightarrow j}$ . In the correspondence limit where $r$ is very large and neither $s$ nor $t$ differ much from $r$, the transition frequencies  $\nu_{sr}$ and $\nu_{rt}$ become equal to the orbital frequencies in the orbits characterized by the values $rh$, $sh$, and $th$ for the action variable $J$. For the harmonic oscillator, the sums in eq.\ (\ref{eq:2.43}) degenerate to a single term each (with $s=r+1$, $t=r-1$), and the transition frequencies $\nu_{sr},\nu_{rt}$ are all equal to the mechanical frequency $\nu_{0}$. In  sec.\ 6.2 we shall explain Van Vleck's derivation of eq.\ (\ref{eq:2.43}) in detail.
    
From the point of view of classical dispersion theory, the terms in the second sum in eq.\ (\ref{eq:2.43}) for polarization, corresponding to transitions to lower states, have no direct physical interpretation. They appear to correspond to oscillators of negative mass (Kramers 1924a, 676; 1924b, 311)! In the early spring of 1924 Van Vleck had already derived an expression for absorption with a structure similar to that of eq.\ (\ref{eq:2.43}) for polarization (see sec.\ 6.3). In the case of absorption, the terms with transitions to lower states are readily recognized as corresponding to ``negative absorption," i.e., to the process of stimulated emission introduced by Einstein.

As we indicated above, there is some disagreement in the historical literature as to who was (or were) responsible for the key move in the construction of the quantum dispersion formula on the basis of the correspondence principle, viz.\ the replacement (\ref{eq:2.33}) of derivatives with respect to the action variable by difference quotients. \citet[p.\ 193]{Jammer 1966} and \citet[Vol.\ 2, p.\ 173]{Mehra Rechenberg} suggest that Kramers got the idea from Born via Heisenberg. \citet[p.\ 222]{Dresden} makes it crystal clear that Kramers found the rule before Born, but allows for the possibility that Born found it independently, as Kramers did not state the rule in his first {\it Nature} note \citep{Kramers 1924a}, the only presentation of the Kramers dispersion formula that Born had seen when he wrote \citep{Born 1924}. Van Vleck certainly discovered the replacement (\ref{eq:2.33}) of derivatives by difference quotients for himself. Since \citet{Van Vleck 1924a} announced the correspondence principle for absorption, which he could not have derived without this rule, in a paper submitted in April 1924, whereas \citep{Born 1924} was not received by {\it Zeitschrift f\"{u}r Physik} until June 1924, Van Vleck clearly could not have taken the rule from Born's paper. Writing to Born later in 1924, Van Vleck sounds slightly annoyed at Born's insinuation that he, Van Vleck, somehow did not realize that one needs to replace derivatives by difference quotients to get from classical to quantum-theoretical expressions. In the letter from which we already quoted in sec.\ 2.4, Born wrote to Van Vleck:
\begin{quotation}
I am sending you my paper ÒOn Quantum MechanicsÓ [Born 1924], which pursues a goal similar to yours. While you limit yourself to the correspondence with high quantum numbers, I conversely aim for rigorous laws for arbitrary quantum numbers.\footnote{Born to Van Vleck, October 24, 1924 (AHQP).} 
\end{quotation}
To which Van Vleck replied:
\begin{quotation}
Thank you for your letter and reprint relating to ``Quantum Dynamics, etc" \ldots
I have read with great interest your important, comprehensive article.  There is, as you say, considerable similarity in the subject matter in your article and mine, especially as regards to dispersion\footnote{Van Vleck seems to be talking here about \citep{Van Vleck 1924b, Van Vleck 1924c}, whereas Born was talking about \citep{Van Vleck 1924a}. Born asked Van Vleck to send him ``an offprint of your extensive calculations." Van Vleck obliged:  ``As you requested, I am sending you under separate cover a reprint of Parts I and II of my computations," presumably \citep{Van Vleck 1924b, Van Vleck 1924c}.}  \ldots As noted in your letter you mention more explicitly than do I the fact that formulas of the quantum theory result from those of the classical theory by replacing a derivative by a difference quotient.  I have stressed the asymptotic connection of the two theories but I think it is clear in the content of my article that in the problems considered the classical and quantum formulas are connected as are derivatives and difference quotients.\footnote{Van Vleck to Born, November 30, 1924 (AHQP).}
\end{quotation}
That Kramers, Van Vleck, and possibly Born independently of one another hit upon the same idea, underscores that  the rule (\ref{eq:2.33}) for replacing derivatives by difference quotients is so natural that it readily comes to mind when one is trying to connect quantum-theoretical expressions to classical ones on the basis of the correspondence principle.

\subsection{Emission and absorption classically and quantum-theoretically in the special case of a simple harmonic oscillator}

Before we present Van Vleck's ``correspondence principle for absorption" (for the special case of a simple harmonic oscillator), we gather some useful results from the classical theory of a charged 
oscillator (of natural frequency $\nu_0$)
coupled to a Maxwellian electromagnetic field. Such an oscillator (i) emits
electromagnetic radiation of frequency $\nu_0$ in the absence of an external field,
(ii) absorbs energy from an applied electromagnetic field of frequency $\nu$, and (iii) undergoes
a net displacement coherent with an applied electromagnetic field (or ``polarization", analyzed above).

The Larmor formula  (\ref{eq:2.39}) gives the power loss due to radiation by our charged harmonic oscillator. The energy loss of the oscillating system
can be ascribed  to a radiative reaction force given by
\begin{equation}
\label{eq:3.1}
  F_{\rm rad} = \frac{2e^{2}}{3c^{3}}\ddot{v} \equiv m\tau \ddot{v},
\end{equation} 
where we shall assume that the characteristic time $\tau$ is very short in comparison to the
mechanical period: $\omega_{0}\tau <<1$, so that radiation damping is very slow on the
time scale of the mechanical oscillations of the system. The equation of motion of the 
oscillator (in the absence of external applied forces) now becomes
\begin{equation}
\label{eq:3.2}
  \dot{v}-\tau \ddot{v} + \omega_{0}^{2} x = 0.
\end{equation}
Now, to a good approximation, the coordinates and velocities of this system are still behaving
as harmonic oscillations of frequency $\omega_{0}$ so we may assume $\ddot{v}\simeq -\omega_{0}^{2}v$
in (\ref{eq:3.2}) and obtain 
\begin{equation}
\label{eq:3.4}
  \ddot{x} + \tau\omega_{0}^{2} \dot{x} + \omega_{0}^{2}x = 0.
\end{equation}
Inserting the {\it Ansatz} $x(t)= De^{-\alpha t}$ into equation (\ref{eq:3.4}), we find:
\begin{equation}
(\alpha^2 - \tau \omega_0^2 \alpha + \omega_0^2) D e^{-\alpha t} = 0.
\end{equation}
Neglecting a term with $\tau^2 \omega_0^4$ (recall that $\omega_0 \tau << 1$, so that $\tau^2 \omega_0^4 << \omega_0^2$),\footnote{Such terms
are treated incorrectly  in any event by the approximation leading to eq.\ (\ref{eq:3.4}).} we can rewrite the expression in parentheses as:
\begin{equation}
(\alpha - \frac{1}{2} \tau \omega_0^2 + i \omega_0)(\alpha - \frac{1}{2} \tau \omega_0^2 - i \omega_0).
\end{equation}
It follows that:
\begin{equation}
\label{eq:3.5}
  \alpha \simeq \frac{1}{2}\tau\omega_{0}^{2}\pm i\omega_{0}\equiv \Gamma/2 \pm i\omega_{0}.
\end{equation}
Thus, we have a solution of the form 
\begin{equation}
\label{eq:3.5a}
   x(t)=De^{-\Gamma t/2}\cos{\omega_0 t},
\end{equation}
from which 
the average rate of oscillator energy loss from the Larmor formula (\ref{eq:2.39}) at small times (i.e., when damping
due to the $e^{-\Gamma t/2}$ factor can be ignored) is easily seen to be 
\begin{equation}
\label{eq:3.6}
   -\frac{dE_{\rm osc}}{dt}= \frac{e^{2}}{3c^{3}}D^{2}\omega_0^{4} = \frac{16\pi^{4}e^{2}}{3c^{3}}D^{2}\nu_0^{4}
   \end{equation}
(where we used that $\dot{v} \simeq \omega_0^2 D$). The constant $\Gamma=\tau\omega_{0}^{2}$ is called the radiative decay constant. We
emphasize again that the preceding discussion presupposes the narrow resonance limit, $\Gamma << \omega_0$. In terms of
$\Gamma$, the basic equation of motion (\ref{eq:3.4}) can be written as
\begin{equation}
\label{eq:3.7}
 \ddot{x} + \Gamma\dot{x} + \omega_{0}^{2}x = 0.
\end{equation} 
  
   Now suppose that our charged oscillator is immersed in an ambient electromagnetic 
   field, characterized by a spectral function (energy density per unit spectral interval)
   $\rho(\nu)$. As we are dealing with one-dimensional oscillators we shall simplify
   the discussion by assuming that only the $x$-component of the electric field is relevant as all the
   oscillators are so aligned. Then (using overbars to denote time averages) the average value of the 
   electromagnetic energy density is (in Gaussian units)
   $(1/4\pi)\bar{\bf{E}}^2=(3/4\pi)\bar{E_{x}}^{2}=\rho(\nu)\Delta\nu$ in the
   frequency interval $(\nu,\nu+\Delta\nu)$. If $E_{x}=E\cos{2\pi\nu t}$ we 
   have $\bar{E_{x}}^{2} = E^{2}/2$ so finally we have
   \begin{equation}
   \label{eq:3.8}
    E^{2}=\frac{8\pi}{3}\rho(\nu)\Delta\nu.
   \end{equation}
   The equation of motion (\ref{eq:3.7}) must be modified to include the coupling to
     the external field (switching back temporarily to angular frequencies, $\omega=2\pi\nu$,
     and using complex notation to encode amplitude and phase information):
     \begin{equation}
     \label{eq:3.9}
       \ddot{x}+\Gamma\dot{x}+\omega_{0}^{2}x= \frac{eE}{m} e^{i\omega t}\equiv F_{\rm app}/m,
       \end{equation}
and the average rate of energy absorption of the oscillator from the ambient field is simply the time
average $<F_{\rm app}\dot{x}>$. This linear second order equation is solved by a sum of 
transients (i.e. solutions of the homogeneous equation: see eq.\ (\ref{eq:3.7})) 
\begin{equation}
  x_{\rm tr}(t) =  De^{-\Gamma t/2}\cos{\omega_0 t},
\end{equation} 
plus the following particular solution
coherent with the applied perturbation
\begin{equation}
\label{eq:3.10}
   x_{\rm coh}(t) = {\rm Re} \;\;\frac{eE}{m}\frac{e^{i\omega t}}{\omega_{0}^{2}-\omega^{2}+i\Gamma\omega},
\end{equation}
so that the desired time average $<F_{\rm app}\dot{x}>=<F_{\rm app}(\dot{x}_{\rm tr}+\dot{x}_{\rm coh})>$ giving the energy absorption rate becomes
\begin{equation}
\label{eq:3.11}
 <F_{\rm app}\dot{x}>=<eE\cos{\omega t} \; \frac{eE}{m} \; {\rm Re} \left( \frac{i\omega e^{i\omega t}}{
 \omega_{0}^{2}-\omega^{2}+i\Gamma\omega} \right)>.
 \end{equation}
 Note that the transient part of the particle coordinate $x_{\rm tr}(t)$ is {\it not} coherent with the
 applied field (we assume $\omega \neq \omega_0$), and therefore does not contribute to the time average
 of the energy absorption. This explains why the amplitude $D$ of the oscillations is absent from the
 final result, which will instead depend only on the specific energy density of the ambient field. In other words,
 even though the charged particle may be executing very large amplitude oscillations $x_{\rm tr}(t)$, the only part
 of the full coordinate $x(t)$ responsible for a nonvanishing average absorption is
 the part of the displacement $x_{\rm coh}(t)$ induced by the applied field, which is proportional to $E$
 and does not involve the amplitude $D$. As we shall see below, the corresponding feature of the quantum
 calculation in the correspondence limit led Van Vleck to the very important realization that the net energy
 absorption involves a {\it difference} in the amount of absorption and stimulated emission as described in Einstein's
 quantum theory of radiation.
 
 Only the cosine part of the complex exponential in eq.\ (\ref{eq:3.11}) will contribute to the
 time average. Using $<\cos^{2}{\omega t}>=1/2$ and eq.\ (\ref{eq:3.8}), we find
 \begin{eqnarray}
 <F_{\rm app}\dot{x}>&=& \frac{e^{2}E^{2}\Gamma}{2m}\frac{\omega^{2}}{(\omega_{0}^{2}-\omega^{2})^{2}+\Gamma^{2}\omega^{2}} \nonumber  \\
 \label{eq:3.12}
  &=& \frac{4\pi e^{2}}{3m}\rho(\frac{\omega}{2\pi})\Gamma\frac{\omega^{2}}{(\omega_{0}^{2}-\omega^{2})^{2}+\Gamma^{2}\omega^{2}}  \frac{1}{2\pi}\Delta\omega
  \end{eqnarray}
for the energy absorption rate due to the ambient field in the frequency interval $(\nu,\nu+\Delta\nu)
=(\omega,\omega+\Delta\omega)$. Since eq.\ (\ref{eq:3.12}) contains the electric field $E$
 squared,  it is apparent that the 
generalization of this linear simple harmonic oscillator result to an arbitrary multiply-periodic system will
require a second-order canonical perturbation theory calculation, which will necessarily be more 
involved than the corresponding classical polarization calculation, which only involves the
electric field to the first order.
In the case of interest, where $\Gamma <<
\omega_0$, the line resonance shape in eq.\ (\ref{eq:3.12}) is highly peaked around the resonance
frequency $\omega_0$, so we may use the distributional limit
\begin{equation}
\label{eq:3.13}
  \frac{\epsilon}{x^{2}+\epsilon^{2}} \rightarrow \pi\delta(x),\;\;\;\epsilon\rightarrow 0
  \end{equation}
 with $x=\omega^{2}-\omega_{0}^{2}$ and $\epsilon=\Gamma\omega$ to execute the integration over $\omega$ in eq.\ (\ref{eq:3.12}) and compute the total absorption rate:
\begin{eqnarray}
 <F_{\rm app}\dot{x}>&\approx& \frac{2e^{2}}{3m}\int \rho(\frac{\omega}{2\pi})\Gamma\frac{\pi}{\Gamma\omega}\omega^{2}\delta(\omega^{2}-\omega_{0}^{2})d\omega  \nonumber \\
\label{eq:3.14}
  &=& \frac{\pi e^{2}}{3m}\rho(\nu_{0}). 
 \end{eqnarray}
This classical result is found in \citep{Planck 1921}  \citep[p.\ 339, note 12]{Van Vleck 1924b})\footnote{Van Vleck probably got the references to \citep{Planck 1921} from \citep{Ladenburg and Reiche 1923}. Both \citep[p.\ 339, note 12; p.\ 340, note 14]{Van Vleck 1924b} and \citep[p.\ 588, note 19; p.\ 591, note 30]{Ladenburg and Reiche 1923} cite ``equations (260) and (159)" and ``section 158" in \citep{Planck 1921}. 
\label{planck}} and
  gives the rate at which a classical charged
   oscillator gains energy  when immersed in an ambient classical
   electromagnetic field. 
   
  In eq.\ (\ref{eq:2.41}) we found the connection in the limit of
  high quantum numbers between
  the Einstein $A$ coefficients and the amplitudes $D_r=2|A^{r}|$ of the mechanical motion
  in the emitting state $r$:
 \begin{eqnarray}
 \label{eq:3.15}
     A_{r\rightarrow s}& \simeq& \frac{16\pi^{4}e^{2}}{3hc^{3}}D_{r}^{2}\nu_{rs}^{3}.
     \end{eqnarray}
From the Einstein
 relation (\ref{eq:2.37}) this  implies a corresponding result for the $B$-coefficients:
 \begin{equation}
 \label{eq:3.16}
   B_{r\rightarrow s}=B_{s\rightarrow r} = \frac{2\pi^{3}e^{2}}{3h^{2}}D_{r}^{2}.
   \end{equation}

 In the $r$-th quantized state of the oscillator, we have $J=rh$
 so from eq.\ (\ref{eq:2.13a}) the corresponding amplitude $D^{\rm qu}_{r}$ of the {\it quantized} motion becomes
 \begin{equation}
 \label{eq:3.17}
   D_{r}^{\rm qu}    = \sqrt{\frac{rh}{2\pi^{2}m\nu_{0}}},
   \end{equation}
   and the quantum result  for the $A$ coefficients in the present case of a linear simple harmonic
   oscillator becomes
   \begin{equation}
   \label{eq:3.18}
   A_{r\rightarrow r-1} = \frac{8\pi^{2}e^{2}\nu_{0}^{2}r}{3mc^{3}},
   \end{equation}
  while the quantum result for the $B$ coeffficients takes the form
   \begin{equation}
   \label{eq:3.19}
     B_{r\rightarrow r-1}=B_{r-1\rightarrow r}=\frac{\pi e^{2}r}{3hm\nu_{0}}.
     \end{equation}

   The Einstein analysis of $A$ and $B$ coefficients makes it
   clear that at the quantum level we must consider what \citet[p.\ 340]{Van Vleck 1924b} calls the
   ``differential absorption rate": the rate of energy absorption of the oscillator in
   state $r$ going to state $r+1$ via (\ref{eq:2.36}) {\it minus} the 
   stimulated emission induced by the ambient
   field and causing the transition $r$ to $r-1$ (the $B$ term in (\ref{eq:2.35})).
    From eq.\ (\ref{eq:3.19}) we therefore have for the differential absorption rate
   of an oscillator in state $r$
   \begin{eqnarray}
     \frac{dE_{\rm net}}{dt}   &=& h\nu_{0}(B_{r\rightarrow r+1}-B_{r\rightarrow r-1})\rho(\nu_{0}) \nonumber \\
       &=& h\nu_{0}(B_{r+1\rightarrow r}-B_{r\rightarrow r-1})\rho(\nu_{0}) \nonumber \\
       &=& h\nu_{0}(r+1-r)\frac{\pi e^{2}}{3hm\nu_{0}}\rho(\nu_{0}) \nonumber \\
       &=& \frac{\pi e^{2}}{3m} \rho(\nu_0),
       \label{eq:3.19a}
    \end{eqnarray}
       which is precisely the classical result (\ref{eq:3.14}).  Note that the dependence on the quantum
       state $r$ (or classically, the amplitude of the motion $D_r$) has {\rm cancelled} in the differential
       absorption rate, corresponding to the lack of coherence discussed previously between the transient
       and impressed motion.
       
Van Vleck derived this result in sec.\ 4 of his paper. He concluded:
   \begin{quotation}
   We thus see that in the limiting case of large quantum numbers, where [eq.\ (\ref{eq:3.19})] is valid, the classical value [in eq.\ (\ref{eq:3.14})] for the rate of absorption of energy is nothing but the differential rate of absorption in the quantum theory. This connection of the classical and quantum differential absorption we shall term the correspondence principle of absorption \citep[p.\ 340]{Van Vleck 1924b}.\footnote{Van Vleck points out that this ``is a purely mathematical consequence of the correspondence principle for emission, which was used  in deriving [eq.\ (\ref{eq:3.19})]" (ibid.). A few pages later, \citet[p.\ 343]{Van Vleck 1924b} notes that he could also have done the reverse, deriving the correspondence principle for emission from that for absorption.}
   \end{quotation}
In sec.\ 5, he generalized the result to arbitrary non-degenerate multiply-periodic systems.  

Van Vleck's correspondence principle for `differential absorption' (i.e., the excess of absorption over stimulated emission) also clarifies the correspondence principle for dispersion. As \citet{Kramers 1924a,Kramers 1924b}
emphasized, the negative terms in the dispersion formula were
difficult to account for on the basis of purely classical concepts---they somehow corresponded
to a {\it negative} value for $e^2/m$ for those virtual oscillators corresponding to transitions
from the initial atomic level to lower energy levels (see sec.\ 3.4). Similar negative contributions
in the case of absorption are physically much more transparent: transitions to higher levels result in a positive absorption of energy from the ambient electromagnetic
field, whereas transitions to lower levels result in energy being returned to the
field. The latter process was therefore known as ``negative absorption" at the time, a term used by both \citet[p.\ 676]{Kramers 1924a}
and \citet[p.\ 338]{Van Vleck 1924b}. Noticing the greater physical transparency of his
correspondence-principle results for absorption, and under the impression that
Kramers' correspondence-principle arguments for the dispersion formula rested only on
a treatment of harmonic oscillators,
Van Vleck  added sections on dispersion to his paper. Sec.\ 6, ``The General Correspondence Principle Basis for KramersÕ Dispersion Formula," was added to the first quantum-theoretical part of the paper; sec.\ 15, ``Computation of Polarization," to the classical part (see the letter from Van Vleck to Kramers of September 1924, quoted in sec.\ 3.4). Van Vleck was thus the first to publish a fully explicit
derivation of the correspondence limit for  polarization
in the context of a general multiply-periodic system.

When Kuhn in his AHQP interview with Van Vleck brought up the paper on the correspondence principle for absorption, Van Vleck said: ``I think that was one of my better papers." ``How did you get into that?," Kuhn wanted to know. Van Vleck told him:
\begin{quotation}
Through a misunderstanding of something Gregory Breit [Van Vleck's colleague in Minnesota at the time] told me. He said that the net absorption was the difference between the fluctuations up and the fluctuations down, referred to some paper of---I think it was (Kretschmann)---but that was an entirely different thing. It was concerned with the fact that under certain phase relations the light did work on the atom and under certain phase relations the atom did work on the light. It was dealing essentially with statistical fluctuations. I misunderstood his remark and proceeded to try and get the differential effect between the absorption up from a given stationary state and a[b]sorption going down.\footnote{P.\ 22 of the transcript of the first session of the AHQP interview with Van Vleck. Van Vleck told this story in somewhat greater detail to Katherine Sopka. He also explained to her why he acknowledged Breit in \citep[p.\ 28]{Van Vleck 1924a} but not in \citep{Van Vleck 1924b, Van Vleck 1924c}: ``As he [Van Vleck] remembers it, he wanted to thank Breit in the latter, but Breit objected on the ground that the phase fluctuations he had in mind were quite different from the difference effect employed by Van Vleck and so, overmodestly, felt no acknowledgment was in order" \citep[p.\ 135, note 184; this note makes no mention of Kretschmann]{Sopka}.}
\end{quotation}
The paper Breit was referring to is presumably \citep{Kretschmann 1921}. In this paper, Erich Kretschmann (1887--1973), a student of Planck better known for his work in general relativity \citep{Kretschmann 1917}, gave a purely classical discussion of the emission and absorption of radiation. What Van Vleck says here about this paper fits with its contents. 

Van Vleck's comments, however, are also very reminiscent of the following passage in 
 \citep{Ladenburg and Reiche 1923}:
\begin{quotation}
\ldots according to Einstein's assumptions the effect of external radiation on a quantum atom corresponds to the effect a classical oscillator experiences from an incident wave. When the frequency of such a wave does not differ much or not at all from the characteristic frequency of the oscillator, the reaction of the oscillator consists in an increase or a decrease of its energy, depending on the difference in phase between the external wave and the motion of the oscillator. In analogy to this, Einstein assumes that the atom in state $i$ has a probability characterized by the factor $b_{ik}$ to make a transition to a higher state $k$ under absorption of the energy $h \nu$ of the incident wave (``positive irradiation") and that the atom in state $k$ has another probability ($b_{ki}$) to return to the state $i$ under the influence of an external wave (``negative irradiation") \citep[p.\ 586]{Ladenburg and Reiche 1923} 
\end{quotation}
As we mentioned in sec.\ 3.3, Ladenburg and Reiche appealed to the correspondence principle to justify their quantum formulae for emission, absorption, and dispersion. Except in the case of emission, however, their arguments were fallacious. We conjecture that this is what inspired Van Vleck to use his expertise in techniques from celestial mechanics---the kind of expertise Ladenburg and Reiche clearly lacked---to derive the correct expressions for emission and absorption merging with classical results in the sense of the correspondence principle.\footnote{As we saw in sec.\ 3.4, Van Vleck's calculations for dispersion were inspired by \citep{Kramers 1924a}.} \citet[p.\ 339, note 13; p.\ 344, note 21]{Van Vleck 1924b} cited Ladenburg and Reiche but gave no indication that their paper was an important source of inspiration for his own. It is not implausible, however, that Van Vleck simply preferred to pass over their badly flawed calculations in silence rather than touting his own clearly superior results. As we mentioned in sec.\ 3.3, one of the problems with the ``correspondence" arguments of Ladenburg and Reiche is that, following \citep{Planck 1921} and in the spirit of the derivation of the $A$ and $B$ coefficients in \citep{Einstein 1917a}, they focus on collections of atoms in thermal equilibrium rather than on individual atoms. What is suggestive of a possible influence of \citep{Ladenburg and Reiche 1923} on \citep{Van Vleck 1924b, Van Vleck 1924c} is that the exact same passages of \citep{Planck 1921} are cited in both papers (see note \ref{planck} above) and that \citet{Van Vleck 1924b} explicitly comments on the issue of many atoms in thermal equilibrium versus single atoms, noting that in Planck's discussion ``no explicit mention is made of the asymptotic connection of the classical absorption and the differential absorption for a single orbit (where thermodynamic equilibrium need not be assumed) which is the primary concern of the present paper" (p.\ 340, note 14).  The topic of a third installment that Van Vleck originally planned to add to his two-part paper also becomes understandable in light of our conjecture about the connection between \citep{Van Vleck 1924b, Van Vleck 1924c} and \citep{Ladenburg and Reiche 1923}. As Van Vleck explained in 1977 (see sec.\ 2.4): ``Part III was to be concerned with the equilibrium between absorption and emission under the Rayleigh-Jeans law" \citep[p.\ 939]{Van Vleck and Huber 1977}. If Ladenburg and Reiche did indeed stimulate Van Vleck's work, however, it is somewhat puzzling that he does not seem to have recognized that the virtual oscillators of BKS, which, as we saw in secs.\ 3.4, 4.1, and 4.2, he consistently attributed to Slater, were essentially just the substitute oscillators of \citep{Ladenburg and Reiche 1923}. We also saw, however, that Van Vleck was hardly alone in associating virtual oscillators with BKS. We thus conclude that it is plausible that Van Vleck was inspired by \citep{Ladenburg and Reiche 1923} to formulate correspondence principles for emission and absorption. For one thing, this would explain why Van Vleck, who had not worked on radiation theory before, turned his attention to the interaction between matter and radiation.

  
\section{Generalization to arbitrary non-degenerate multiply-periodic systems}

\subsection{The correspondence principle for absorption}    
        
    The primary result of \citep{Van Vleck 1924b,Van Vleck 1924c}
    was an extension of eq.\ (\ref{eq:3.19a}) to an arbitrary non-degenerate multiply-periodic
    system of a single particle in three dimensions, and the demonstration that the quantum-differential
    absorption coincides with this more general result in the correspondence limit. Before giving Van Vleck's
    result we recall some basic features of multiply-periodic systems, which we shall in any event need in
    the next subsection when a completely explicit derivation (following, with minor notational
    changes, the one laid out by Van Vleck)  of the corresponding formula for polarization
    will be provided.
    
     The transition from one-dimensional periodic (and harmonic) systems such as the linear simple harmonic
     oscillator to three-dimensional
 multiply-periodic ones is fairly straightforward.  Apart from the obvious need to introduce vector quantities, there are only two 
 significant additional features. First, there is the appearance of multiple overtones in the general multiply-periodic
 expansion (so that the multiplicity variables in the analogue of eq.\ (\ref{eq:2.20}) take arbitrary
 positive and negative integral values, not just $\pm 1$). Second, the mechanical frequencies
 $\nu_1,\nu_2,\nu_3$ (with $\nu_i = \partial H_{0}/ \partial J_{i}$) of the separated coordinates are now in general functions of the
 amplitude of the classical path, which is to say, of the action variables $J_{i}$ (with $i=1,2,3$). We
 assume as before that the imposed electric field is in the $X$-direction so the $x$-coordinate
 of our electron is the relevant one for computing the induced coherent polarization, and in
 analogy to eq.\  (\ref{eq:2.20}) we now have
 \begin{equation}
 \label{eq:3.20}
   x(t) = \sum_{\vec{\tau}}A_{\vec{\tau}} e^{2\pi i \vec{\tau}\cdot\vec{w}},
  \end{equation}
  where in the absence of the external field the angle variables $\vec{w}=(w_1,w_2,w_3)=
  (\nu_1,\nu_2,\nu_3)t\equiv\vec{\nu}t$ and $\vec{\tau}=(\tau_1,\tau_2,\tau_3)$ with $\tau_{i}$ taking
  on all (positive and negative) integer values. It will be useful to write eq.\ (\ref{eq:3.20})
  in an alternative purely real form, as a cosine expansion:
  \begin{equation}
  \label{eq:3.21}
    x(t) = \sum_{\vec{\tau},\vec{\tau}\cdot\vec{\nu}>0} X_{\vec{\tau}}\cos{(2\pi\vec{\tau}\cdot\vec{\nu}t)}.
    \end{equation} 
 The complex amplitudes $A_{\vec{\tau}}$ satisfy the conjugacy condition
 $A_{\vec{\tau}}=A^{*}_{-\vec{\tau}}$ to ensure that $x(t)$ is real and we have the relation $X_{\vec{\tau}}^{2}=4A_{\vec{\tau}}A_{-\vec{\tau}}$.\footnote{Cf.\ eqs.\ (\ref{eq:2.20})--(\ref{eq:2.20a}) and note \ref{amp} in sec.\ 6.1.}
 
   As before (cf.\ eq.\ (\ref{eq:2.19})), the full Hamiltonian has the form
 \begin{equation}
 \label{eq:3.22}
   H = H_0 +eEx(t)\cos{2\pi\nu t}.
   \end{equation}
 The subscripted mechanical frequencies $\nu_{i}$ with $i=1,2,3$ (comprising the
   vector $\vec{\nu}$) must be distinguished from the single
   frequency $\nu$ (unsubscripted) corresponding to the applied field.
   
   With these notations, Van Vleck's result for the absorption rate becomes \citep[p.\ 342, eq.\ (16)]{Van Vleck 1924b}:
   \begin{equation}
   \label{eq:3.22a}
   \frac{dE_{\rm net}}{dt}=\frac{2}{3}\pi^{3}e^{2}\left[\rho(\vec{\tau}\cdot\vec{\nu})\tau_{k}\frac{\partial G_{\tau}}{\partial J_{k}}
   +\rho^{\prime}(\vec{\tau}\cdot\vec{\nu})G_{\tau}\tau_{k}\frac{\partial}{\partial J_{k}}(\vec{\tau}\cdot\vec{\nu})\right].
   \end{equation}
where $\rho^{\prime}\equiv\partial\rho/\partial\nu$ and where summation over $k = (1, 2, 3)$ is implied and where $G_{\tau}\equiv \vec{\tau}\cdot\vec{\nu}D_{\vec{\tau}}^{2}$ with $D_{\vec{\tau}}^{2}\equiv X_{\vec{\tau}}^{2}+Y_{\vec{\tau}}^{2}
+Z_{\vec{\tau}}^{2}$.
  In the special case of the harmonic oscillator, the term with $\rho^{\prime}$, the derivative of the spectral function, vanishes as there is only a single
  mechanical frequency $\nu=\nu_0$, which is independent of the action variable $J$. In the first term, we get simply
  \begin{equation}
  \label{eq:3.22b}
   \frac{dE_{\rm net}}{dt}=\frac{2}{3}\pi^{3}e^{2}\rho(\nu_0)\frac{\partial}{\partial J}(\nu_{0}D^{2}).
   \end{equation}
   Using eq.\ (\ref{eq:2.13a}), $D=\sqrt{J/m\pi\omega_{0}}$, for the amplitude, we recover the previous result, eq.\ (\ref{eq:3.19a}).
   
    Eq.\  (\ref{eq:3.22a}) is the product of a highly nontrivial application of canonical
    perturbation techniques, where quantities of {\it second} order in the applied field need
    to be properly evaluated (cf. discussion following eq.\ (\ref{eq:3.12}) above). The
    polarization calculation presented in full in the next section only involves canonical
    perturbation theory to first order. For the absorption calculation, the
    variation in the action variables $\Delta J_{k}$ in particular is needed to second order, and the integration of the result obtained for a monochromatic incident field
    needed to pass to the case of continuous radiation specified by an arbitrary
    spectral function $\rho(\nu)$ requires considerable care.
    
    Slater also tried his hand at this calculation, as can be inferred from a letter from Kramers to Van Vleck, from which we already quoted in sec.\ 3.4. Kramers wrote:
\begin{quotation}
Slater had, on my request, made the same calculation, and he stated that the classical mean-absorption formula gave the right result in the limit of high quantum numbers. I did, however, not see his formula, and am not quite sure that he had not forgotten the term with $\partial \rho / \partial \nu$, without which the thing is not complete of course.\footnote{Kramers to Van Vleck, November 11, 1924 (AHQP).}
 \end{quotation}
 Van Vleck clearly remembered this point almost forty years later. Talking to Kuhn about his 1924 absorption papers, he mentioned: ``I got the term in partial rho with respect to nu. I'm very proud of the fact that I picked that one up \ldots Slater, at Kramers' suggestion I guess, made a completely parallel calculation in Copenhagen which he never published."\footnote{P.\ 22 of the transcript of the first session of the AHQP interview with Van Vleck.} 
 

\subsection{The correspondence principle for polarization}

 In this section we retrace the derivation given by  \citet{Van Vleck 1924c} of the
 classical polarization formula for a general non-degenerate multiply-periodic system (with a single
 electron) in three dimensions. We remind the reader that this result is by no means new
 to Van Vleck, nor, for that matter, to  Born or Kramers, who also produced derivations of the same result at around
 this time, using slightly different versions of canonical perturbation theory (cf.\ our comments at the end of sec.\ 6.1).
 The formula obtained is basically identical to a formula originally derived in celestial mechanics to compute the
 perturbation in the orbits of the inner planets due to the outer ones. As we saw in sec.\  3.2, Epstein had been the first to use the relevant techniques from celestial mechanics in the context of the old quantum theory. As Van Vleck reminded Slater: ``The classical formula analysis to the Kramer[s] formula appears to be first ca[lc]ulated by Epstein [1922c]."\footnote{Van Vleck to Slater, December 15, 1924 (AHQP).}
  
   The derivation is basically a straightforward generalization of the derivation of sec.\  5.1 for the special case of a charged simple harmonic
 oscillator in an electromagnetic field (see eqs.\ (\ref{eq:2.22})--(\ref{eq:2.31})). The first-order perturbation in the coordinate $x(t)$ (the direction of the electric field in the incident electromagnetic wave)
corresponding to the shifts $(\Delta J_{l}, \Delta w_{l})$ in the action-angle
   variables  is given by the three-dimensional version of eq.\ (\ref{eq:2.22}):
  \begin{equation}
  \label{eq:3.22c}
     \Delta x = \frac{\partial x}{\partial J_{l}}\Delta J_{l} + \frac{\partial x}{\partial w_{l}}\Delta w_{l}.
  \end{equation} 
As in sec.\ 5.1, 
  we imagine that the external field is switched on at time zero, so that the shifts $(\Delta J_l, \; \Delta w_l)$ are the integrals of their time derivatives from $0$ to $t$. In
   analogy with eq.\ (\ref{eq:2.25}) and using eq.\ (\ref{eq:3.20}) for $x(t)$, we can immediately write down the equation for
    $\Delta J_l$ to first order in $E$:
   \begin{equation}
   \label{eq:3.23}
   \Delta J_{l} = \int_0^t \dot{J_{l}}dt = \frac{eE}{2}\sum_{ \vec{\tau}} \tau_{l}A_{\vec{\tau}}\left\{\frac{1-e^{2\pi i(\vec{\tau}\cdot\vec{\nu}+\nu)t}}{\vec{\tau}\cdot\vec{\nu}+\nu}+(\nu\rightarrow -\nu)\right\}.
   \end{equation}
   All the terms  inside
   the summation can be taken to zeroth order in the applied field.
 The computation of the first-order shifts $\Delta w_{l}$ is a little more involved as
  new terms, not present in the harmonic-oscillator case, enter (cf.\ note \ref{nu_0(J)}).  
 The Hamilton equation for $\dot{w_{l}}$ for the full Hamiltonian eq.\ (\ref{eq:3.22}) 
  is (cf.\ eq.\ (\ref{eq:2.26})):
  \begin{equation}
  \label{eq:3.24}
   \dot{w_{l}} = \nu_{l} +\frac{eE}{2}\sum_{\vec{\tau}}\frac{\partial A_{\vec{\tau}}}{\partial J_{l}}\left\{
   e^{2\pi i(\vec{\tau}\cdot\vec{\nu}+\nu)t}+(\nu\rightarrow -\nu)\right\}.
   \end{equation}
   Both terms in eq.\ (\ref{eq:3.24}) contribute to the first-order deviation $\Delta \dot{w}_{l}$ from the value of $\nu_l$ for the unperturbed system. Since $\nu_l$ depends on $J_k$, there will be a term $(\partial \nu_l / \partial J_k) \Delta J_k$ (cf.\ note \ref{nu_0(J)}). The second term is just the generalization of the corresponding term in eq.\ (\ref{eq:2.26}). Hence, we get:
   \begin{equation}
 \label{eq:3.25}
    \Delta\dot{w_{l}}   =   \frac{\partial \nu_{l}}{\partial J_{k}}\Delta J_{k}+\frac{eE}{2}\sum_{\vec{\tau}}\frac{\partial A_{\vec{\tau}}}{\partial J_{l}}\left\{
   e^{2\pi i(\vec{\tau}\cdot\vec{\nu}+\nu)t}+(\nu\rightarrow -\nu)\right\}.
 \end{equation}
Upon substitution of eq.\  (\ref{eq:3.23}) for $\Delta J_k$ this turns into
 \begin{eqnarray}
 \label{eq:3.26}
  \Delta\dot{w_{l}} & = & \frac{eE}{2}\sum_{\vec{\tau}}\left\{ \frac{\partial A_{\vec{\tau}}}{\partial J_{l}}
  e^{2\pi i(\vec{\tau}\cdot\vec{\nu}+\nu)t}+\tau_{k}\frac{\partial\nu_{l}}{\partial J_{k}}A_{\vec{\tau}}
  \frac{1-e^{2\pi i(\vec{\tau}\cdot\vec{\nu}+\nu)t}}{\vec{\tau}\cdot\vec{\nu}+\nu} \right\}  \\ & &  \;\;\;\;\;\; + \;\; (\nu\rightarrow -\nu). \nonumber
  \end{eqnarray}
      Integrating eq.\ (\ref{eq:3.26}), we find
  \begin{eqnarray}
  \label{eq:3.27}
   \Delta w_{l} & = & \frac{eE}{4\pi}\sum_{\vec{\tau}}\left\{i\frac{\partial A_{\vec{\tau}}}{\partial J_{l}}
   \frac{1-e^{2\pi i(\vec{\tau}\cdot\vec{\nu}+\nu)t}}{\vec{\tau}\cdot\vec{\nu}+\nu}  \right. \\ & & \nonumber \left. \;\;\;\;
   +\tau_{k}\frac{\partial\nu_{l}}{\partial J_{k}}A_{\vec{\tau}}\frac{2\pi(\vec{\tau}\cdot\vec{\nu}+\nu)t
   -i(1-e^{2\pi i(\vec{\tau}\cdot\vec{\nu}+\nu)t})}{(\vec{\tau}\cdot\vec{\nu}+\nu)^{2}} \right\} 
   + (\nu\rightarrow -\nu).
 \end{eqnarray}
   
  Inserting eq.\ (\ref{eq:3.20}) into eq.\ (\ref{eq:3.22c}), we arrive at
   \begin{equation}
   \label{eq:3.28}
   \Delta x(t) = \sum_{\vec{\tau}^{\prime}}\left(\frac{\partial A_{\vec{\tau}^{\prime}}}{\partial J_{l}}\Delta J_{l}
   +2\pi iA_{\vec{\tau}^{\prime}}\tau_{l}^{\prime}\Delta w_{l}\right)e^{2\pi i\vec{\tau}^{\prime}\cdot\vec{\nu}t}.
   \end{equation}
Inserting eqs.\ (\ref{eq:3.23}) and (\ref{eq:3.27}) for $\Delta J_l$ and $\Delta w_l$, respectively, into this expression, we obtain 
 \begin{eqnarray}
   \Delta x(t) &=& \frac{eE}{2}\sum_{\vec{\tau},\vec{\tau}^{\prime}}
   \left\{\tau_{l}\frac{\partial A_{\vec{\tau}^{\prime}}}{\partial J_{l}}A_{\vec{\tau}}\frac{1-e^{2\pi i(\vec{\tau}\cdot\vec{\nu}+\nu)t}}{\vec{\tau}\cdot\vec{\nu}+\nu}-\tau_{l}^{\prime}\frac{\partial A_{\vec{\tau}}}{\partial J_{l}}A_{\vec{\tau}^{\prime}}\frac{1-e^{2\pi i(\vec{\tau}\cdot\vec{\nu}+\nu)t}}{\vec{\tau}\cdot\vec{\nu}+\nu} \right. \nonumber \\
   & &  \;\;\;\;\;\;\;\;\;\;\;\;\;\; + A_{\vec{\tau}}A_{\vec{\tau}^{\prime}}\tau_{k}\frac{\partial\nu_{l}}{\partial J_{k}}\tau_{l}^{\prime}
   \frac{2\pi i(\vec{\tau}\cdot\vec{\nu}+\nu)t+1-e^{2\pi i(\vec{\tau}\cdot\vec{\nu}+\nu)t}}{(\vec{\tau}\cdot\vec{\nu}+\nu)^{2}} \nonumber \\
   \label{eq:3.29}
   & & \;\;\;\;\;\;\;\;\;\;\;\;\;\; 
   + \left. (\nu\rightarrow -\nu) \frac{}{} \right\} e^{2\pi i\vec{\tau}^{\prime}\cdot\vec{\nu}t}.
   \end{eqnarray}
 As in sec.\ 5.1, we are only interested in the coherent contribution to the polarization,
  so we drop all terms in eq.\ (\ref{eq:3.29}) whose time dependence is
 not precisely
 $e^{\pm 2\pi i\nu t}$  and find, writing for convenience $\tau_{k} (\partial/\partial J_{k})\equiv
 \vec{\tau}\cdot\vec{\nabla}_{J}$,
 \begin{eqnarray}
 \label{eq:3.30}
  \Delta x_{\rm coh} & = & \frac{eE}{2}\sum_{\vec{\tau}} \left\{-\vec{\tau}\cdot\vec{\nabla}_{J}(A_{\vec{\tau}}A_{-\vec{\tau}})\frac{e^{2\pi i\nu t}}{\vec{\tau}\cdot\vec{\nu}+\nu} \right. \\ & & \nonumber \left.
 \;\;\;\;\;\;\;\;\;\;\;\; + \;\; A_{\vec{\tau}}A_{-\vec{\tau}}\vec{\tau}\cdot\vec{\nabla}_{J}(\vec{\tau}\cdot\vec{\nu})\frac{e^{2\pi i\nu t}}{(\vec{\tau}\cdot\vec{\nu}+\nu)^{2}}\right\} \; + \;(\nu\rightarrow -\nu).
  \end{eqnarray}
 Note that the coherent contribution derives from terms in which $\vec{\tau}^{\prime}=-\vec{\tau}$,
 as otherwise the uncancelled overtones from the mechanical system would shift the spectral
 line (as in Raman scattering). 
 Essentially the only additional physics of \citep{Kramers and Heisenberg 1925} in comparison to 
 \citep{Van Vleck 1924b, Van Vleck 1924c} 
 is a detailed examination of such terms, predicted earlier by \citet{Smekal 1923}. 
 The terms in eq.\ (\ref{eq:3.30}) involving $\sin{2\pi\nu t}$ vanish, as can be seen with the help of the identities
 \begin{eqnarray*}
 \sum_{\vec{\tau}}\tau_{j} \left(\frac{1}{\vec{\tau}\cdot\vec{\nu}+\nu}-\frac{1}{\vec{\tau}\cdot\vec{\nu}-\nu}\right)
 \cdot ({\rm even\; function\; of\;} \vec{\tau}) &=&0  \\
 \sum_{\vec{\tau}}\tau_{j}\tau_{k}\left(\frac{1}{(\vec{\tau}\cdot\vec{\nu}+\nu)^{2}}-\frac{1}{(\vec{\tau}\cdot\vec{\nu}-\nu)^{2}}\right) \cdot ({\rm even\; function\; of\;} \vec{\tau}) &=&0. 
 \end{eqnarray*}
 Thus eq.\ (\ref{eq:3.30}) simplifies to
 \begin{eqnarray}
  \Delta x_{\rm coh} &=& -\frac{eE}{2}\cos{2\pi\nu t}\sum_{\vec{\tau}}\left\{\vec{\tau}\cdot\vec{\nabla}_{J}(\frac{A_{\vec{\tau}}A_{-\vec{\tau}}}{\vec{\tau}\cdot\vec{\nu}+\nu})+(\nu\rightarrow -\nu)\right\}  \nonumber 
  \\
  \label{eq:3.31}
   &=& -eE\cos{2\pi\nu t}\sum_{\vec{\tau}}\vec{\tau}\cdot\vec{\nabla}_{J}\left(\frac{\vec{\tau}\cdot\vec{\nu}A_{\vec{\tau}}A_{-\vec{\tau}}}{(\vec{\tau}\cdot\vec{\nu})^{2}-\nu^{2}}\right).
   \end{eqnarray}
With the replacement $X_{\vec{\tau}}^{2}=4A_{\vec{\tau}}A_{-\vec{\tau}}$, we may go over to 
the cosine form of the expansion in eq.\ (\ref{eq:3.31}) (cf.\ eqs.\ (\ref{eq:3.20})--(\ref{eq:3.21})), summing over only positive values
of $\vec{\tau}\cdot\vec{\nu}$ (with a factor of 2):
\begin{equation}
\label{eq:3.32}
 \Delta x_{\rm coh} = -\frac{eE}{2}\cos{2\pi\nu t}\sum_{\vec{\tau},\vec{\tau}\cdot\vec{\nu}>0}
 \vec{\tau}\cdot\vec{\nabla}_{J}\left(\frac{\vec{\tau}\cdot\vec{\nu}X_{\vec{\tau}}^{2}}{(\vec{\tau}\cdot\vec{\nu})^{2}-\nu^{2}}\right).
 \end{equation}
 This is the generalization of eq.\ (\ref{eq:2.31}) for the harmonic oscillator.
 
   Finally, we obtain the polarization by multiplying the displacement
 by $N_{r}$, the number of electrons per unit
volume (the subscript $r$ refers to the fact that we shall shortly consider only
electrons in a particular quantum state $r$), and by $-e$ for the electron charge
 \begin{equation}
\label{eq:3.33}
 P = N_{r}\frac{e^{2}}{2}E\cos{2\pi\nu t}\sum_{\vec{\tau}\cdot\vec{\nu}>0}
\vec{\tau}\cdot\vec{\nabla}_{J}\left(\frac{\vec{\tau}\cdot\vec{\nu}X_{\vec{\tau}}^{2}}{(\vec{\tau}\cdot\vec{\nu})^{2}-\nu^{2}}\right)
 \end{equation}
which is eq.\ (41) in (Van Vleck, 1924c; in Van Vleck's notation, $\vec{\tau}\cdot\vec{\nu}$ is written
$\omega_{\tau}$) and equivalent to eq.\ 2$^{*}$ in \citep{Kramers 1924b} (see eq.\ (\ref{Kramers}) above).
 
 The equivalence of eq.\ (\ref{eq:3.33}) to the Kramers dispersion formula  (\ref{eq:2.43}) in the correspondence
 limit is sketched in \citep{Kramers 1924b} and fully explained in sec.\ 6 of \citep{Van Vleck 1924b}.\footnote{Cf. Van Vleck to Kramers, September 22, 1924 (AHQP), quoted in sec.\ 3.4.} Here we
 follow the latter. So we begin with eq.\ (\ref{eq:2.43}) for the polarization of a quantized system
 in state $r$, without the factor of 3 corresponding to the assumption that
 all oscillators be aligned with the applied field (rather than randomly in 3 dimensional space),
 and writing $N_{r}$ instead of  $n_{\rm osc}$:
 \begin{equation}
 \label{eq:3.34}
     P_{r} = \frac{N_{r}c^{3}}{32\pi^{4}}E\cos{2\pi\nu t}\left(\sum_{s}\frac{A_{s\rightarrow r}}{\nu_{sr}^{2}(\nu_{sr}^{2}-\nu^{2})}-\sum_{t}\frac{A_{r\rightarrow t}}{\nu_{rt}^{2}(\nu_{rt}^{2}-\nu^{2})}\right).
 \end{equation}
The sums over $s$ (resp. $t$) refer to states higher (resp. lower) in energy than the fixed state $r$
under consideration. In the correspondence limit, we take the state $r$ to correspond to very
high quantum numbers $(n_1,n_2,n_3)$. The states $s,t$ are associated to the central state $r$ in symmetrical pairs:
\begin{eqnarray}
 s&\rightarrow&(n_1+\tau_1,n_2+\tau_2,n_3+\tau_3), \nonumber \\
 \label{eq:3.35}
 r&\rightarrow&(n_1,n_2,n_3),\\
 t&\rightarrow&(n_1-\tau_1,n_2-\tau_2,n_3-\tau_3), \nonumber
 \end{eqnarray}
 with $\vec{\tau}\cdot\vec{\nu}>0$ so that the states $s$ (resp. $t$) do indeed correspond to higher (resp. lower)
 energy states. Furthermore, we assume that $\vec{\tau}\cdot\vec{\nu}<<\vec{n}\cdot\vec{\nu}$
 so that the transitions $s\rightarrow r\rightarrow t$ correspond to very slight changes in the
 classical orbitals (and differences approximate well to derivatives). 
 The Bohr-Sommerfeld quantization condition (1) associates action
 values $J_{i}=n_{i}h$ with a given quantized state, so the formal correspondence principle
 becomes (cf.\ eq.\ (\ref{eq:2.33}) in sec.\ 6.2):
 \begin{equation}
 \label{eq:3.36}
   \delta_{\vec{\tau}}F(\vec{n})\equiv F(\vec{n})-F(\vec{n}-\vec{\tau})\rightarrow h\vec{\tau}\cdot\vec{\nabla}_{J}F.
   \end{equation}
 In this notation,  formula (\ref{eq:3.34}) the polarization can be written as
 \begin{equation}
 \label{eq:3.37}
 P_{r} =\frac{N_{r}c^{3}}{32\pi^{4}}E\cos{2\pi\nu t}\sum_{\vec{\tau}}\delta_{\vec{\tau}}
 \left(\frac{A_{s\rightarrow r}}{\nu_{sr}^{2}(\nu_{sr}^{2}-\nu^{2})}\right), 
 \end{equation}
 with $A_{s\rightarrow r}$ given by Van Vleck's ``correspondence principle for emission" (see eq.\ (\ref{eq:3.15}) and eq.\ (\ref{eq:2.41}))
 \begin{equation}
 \label{eq:3.38}
  A_{s\rightarrow r}=\frac{16\pi^{4}e^{2}}{3hc^{3}}D_{s}^{2}\nu_{sr}^{3},
  \end{equation}
  where $D_{s}^{2}=(X_{\vec{\tau}}^{(s)})^{2}+(Y_{\vec{\tau}}^{(s)})^{2}+(Z_{\vec{\tau}}^{(s)})^{2}$
  is the full vector amplitude squared for the Fourier component of the classical path responsible
  for the transition $\vec{n}+\vec{\tau}\rightarrow \vec{n}$. Substituting eqs.\ (\ref{eq:3.36}) and (\ref{eq:3.38}) into
  eq.\ (\ref{eq:3.37}) and replacing the difference frequency $\nu_{sr}$ by its classical counterpart $\vec{\tau} \cdot \vec{\nu}$, we obtain, :
  \begin{eqnarray}
  P_{r} &=& N_{r}E\cos{2\pi\nu t}
  \frac{c^{3}}{32\pi^{4}}
  \frac{16\pi^{4}e^{2}}{3hc^{3}}h
\sum_{\vec{\tau}\cdot\vec{\nu}>0}\vec{\tau}\cdot\vec{\nabla}_{J}\left(\frac{\vec{\tau}\cdot\vec{\nu}D_{s}^{2}}{(\vec{\tau}\cdot\vec{\nu})^{2}-\nu^{2}}\right)  \nonumber \\
    \label{eq:3.39}
     &=& N_{r}\frac{e^{2}}{2}E\cos{2\pi\nu t}\sum_{\vec{\tau}\cdot\vec{\nu}>0}
     \vec{\tau}\cdot\vec{\nabla}_{J}\left(\frac{\vec{\tau}\cdot\vec{\nu}\frac{1}{3}D_{s}^{2}}{(\vec{\tau}\cdot\vec{\nu})^{2}-\nu^{2}}\right).
     \end{eqnarray}
   With the replacement $\frac{1}{3}D_{s}^{2}\rightarrow X_{\vec{\tau}}^{2}$ appropriate for randomly
   oriented atoms, eq.\ (\ref{eq:3.39}) becomes identical to the classical formula (\ref{eq:3.33}). This shows that the Kramers dispersion formula (\ref{eq:3.34}) does indeed merge with the classical result in the limit of high quantum numbers, as Van Vleck set out to demonstrate.


\section{Derivation of the formulae for dispersion, emission, and absorption in modern quantum mechanics}
 
Describing the impact of the new quantum mechanics on dispersion theory, Van Vleck wrote in 1928:
\begin{quotation}
Dispersion was particularly bothersome in the old quantum theory, which could never explain why the resonance frequencies in dispersion were experimentally the spectroscopic frequencies given by the Bohr frequency condition rather than the altogether different frequencies of motion in orbits constituting the stationary states [cf.\ our discussion in the introduction of sec.\ 3]. The new mechanics, however, yields the Kramers dispersion formula, previously derived semi-empirically from the correspondence principle \ldots As the result of the masterful treatment by Dirac [1927],  a mechanism has at last to a certain extent been found for the previously so mysterious quantum jumps between stationary states \ldots Dirac's work brings out nicely the parallelism between matter and radiation, and their corpuscular and wave aspects, which are complementary rather than contradictory \citep[pp.\ 494--495]{Van Vleck 1928a}.
\end{quotation} 
That same year, in the first installment of what would turn out to be an eight-part paper entitled ``Investigations of anomalous dispersion in excited gases," Ladenburg  likewise provided a brief synopsis of recent developments in dispersion theory:
\begin{quotation}
The first successful treatment of dispersion phenomena on the basis of Bohr's atomic theory implicitly contained the assumption that the orbital frequencies of the Bohr electrons are the special values at which dispersion changes sign.\footnote{At this point, Ladenburg refers to the papers by Sommerfeld, Debye, and Davisson and the criticism of them by Bohr and Epstein that we discussed in sec.\ 3.2.} In contrast to this, the point of departure of the {\it newer} development of dispersion theory is the empirical fact that not the orbital frequencies of the electrons but the frequencies, observable in emission and absorption, of ``quantum jumps,"  i.e., spectral lines, are the singular values of anomalous dispersion. These correspond to the characteristic frequencies of quasi-elastically bound electrons in the classical electron theory [discussed in sec.\ 3.1]. Tying together the notions of this theory with Bohr's atomic theory has taught us that the ``strength" of the dispersion or of the ``substitute oscillators," which at Bohr's suggestion were introduced as carriers of the scattered radiation needed for dispersion, is determined in non-classical fashion by the ``strength," i.e., the probability of quantum jumps[,] and by the density of atoms in the ``lower" atomic state involved in such quantum jumps.\footnote{At this point, Ladenburg refers to his own work, Bohr's favorable reaction to it, and his subsequent work with Reiche, all discussed in sec.\ 3.3.} H.\,A.\,Kramers then showed,\footnote{At this point, Ladenburg refers to Kramers' two {\it Nature} notes and to the Kramers-Heisenberg paper discussed in sec.\ 3.4.} through correspondence considerations, that the dispersion formula obtained by the author [cf.\ eq.\ (8) in sec.\ 3.3] only holds exactly in the case of non-excited or meta-stable atoms; in the case of excited non-meta-stable atoms, which can also make spontaneous transitions to states of lower energy, this formula is incomplete and has to be supplemented by terms of ``negative dispersion," which correspond to the ``negative absorption" [i.e., stimulated emission] of the radiation theory of Planck and Einstein. Thus originated the ``quantum-theoretical dispersion formula" [cf.\ eq.\ (9) in sec.\ 3.4] which has finally been given a fully consistent foundation in quantum mechanics and wave mechanics;\footnote{At this point, Ladenburg refers to the treatments of dispersion in \citep{dreimaenner}, \citep{Schroedinger 1926}, and \citep{Dirac 1927}. For discussion of Schr\"odinger's wave-mechanical treatment of dispersion, see \citep[Vol.\ 5, pp.\ 789--796]{Mehra Rechenberg}.} this new quantum theory completely avoids concepts like orbital frequencies of electrons in stationary states, and one of its points of departure was precisely the quantum-theoretical interpretation of dispersion phenomena mentioned above \citep[pp.\ 15--16]{Ladenburg 1928}
\end{quotation}
Rather than pursuing the history of dispersion post-{\it Umdeutung}, we shall present
our own modern derivations of quantum formulae for dispersion (sec.\ 7.1), (spontaneous) emission (sec.\ 7.2), and absorption (sec.\ 7.3). Seeing how modern quantum mechanics sanctions the formulae found by Kramers, Van Vleck and others in the old quantum theory on the basis of Einstein's quantum theory of radiation and Bohr's correspondence principle will illuminate various aspects of the relation between the old and the new theory. 

First, we show how the orchestra of virtual oscillators of pre-{\it Umdeutung} dispersion theory survives in the guise of a sum over matrix elements of the position operator. Second, we show how the diagonal matrix elements of the fundamental commutation relation for position and momentum, $[X,P] = i \hbar$, are given by the high-frequency limit of the Kramers dispersion formula, a formula known as the Thomas-Kuhn(-Reiche) sum rule \citep{Thomas 1925, Kuhn 1925, Reiche and Thomas 1925}. This formula replaces the Bohr-Sommerfeld condition as the fundamental quantization condition in the {\it Umdeutung} paper (see sec.\ 3.5). Heisenberg obtained the sum rule by applying the procedure introduced in the {\it Umdeutung} paper for translating classical quantities into quantum-theoretical ones to (a derivative of) the Bohr-Sommerfeld quantization condition. He then showed that the sum rule also obtains by comparing the high-frequency limit of the Kramers dispersion formula with the polarization of a charged harmonic oscillator in the limit where $\nu >> \nu_0$ (see our eq.\ (\ref{eq:2.32a})). In hindsight, we can see clearly in the {\it Umdeutung} paper how close Heisenberg came to recognizing the presence of the
commutation relation between position and momentum in the sum rule serving as his quantization condition. As he told Kuhn:
\begin{quotation}
I had written down, as the quantization rule the Thomas-Kuhn sum rule, but I had not recognized that this was just pq minus qp. That I had not seen.\footnote{P.\ 12 of the transcript of session 5 of the AHQP interview with Heisenberg. See also p.\ 9 of the transcript of session 7. Cf.\ our discussion in sec.\ 3.5.}
\end{quotation}
That he did not take this step is probably due to two important obstacles, one conceptual,
the other technical. The conceptual framework of the entire {\it Umdeutung} paper is Lagrangian
(as opposed to Hamiltonian):
the essential problem is to find a quantum-theoretical reinterpretation of the classical
concepts of position $x(t)$ and {\it velocity} $\dot{x}(t)$ of a particle. Indeed, the conventional
symbol for momentum, $p$, appears only {\it once} in the entire paper, in the statement of the
Bohr-Sommerfeld quantization condition (eq. (12) in the paper). From this point on, $p$ is replaced
everywhere by $m\dot{x}(t)$. The canonical connection between position and momentum (so central, ironically, to the canonical perturbation theory that led to the dispersion 
formula in the first place\footnote{Of course, it was also central to \citep{Dirac 1925}.}) seems simply to have vanished from Heisenberg's thinking
at this point. The other, technical, obstacle was 
 an inconvenient division of the sum over quantum states in the sum rule, which, though  very natural from the point of dispersion theory, obscured its connection to a commutator, as we shall see below.

It will also become clear in the course of our modern derivation that the Kramers dispersion formula is an even more general result in modern quantum mechanics than it was in the old quantum theory. In the old quantum theory, it held for any non-degenerate multiply-periodic system with an unperturbed Hamiltonian such that the unperturbed motion can be solved in action-angle variables. In modern quantum mechanics, the result holds for any system with a Hermitian Hamilton operator such that the unperturbed part has a spectrum that is at least partially discrete. This helps to explain why the Kramers dispersion formula carries over completely intact from the old quantum theory to modern quantum mechanics.


\subsection{Dispersion}

  In this subsection, we derive the Kramers dispersion formula in time-dependent perturbation
  theory. We then examine the high-frequency limit of this formula and discuss the role it played in \citep{Heisenberg 1925c} paper as the fundamental quantization condition replacing the Bohr-Sommerfeld condition.  
 
   We consider a quantized charged system (valence electron) with states labeled by discrete
   indices $r,s,t,...$, and with the Hamilton operator
   \begin{equation}
   \label{eq:quantHam}
     H = H_0 + V(t) = H_0 + e E x\cos{\omega t}.
     \end{equation}
     We want to calculate the first-order perturbation (in the electric field $E$) in the
     expectation value of the electron position in a particular state $|r,t>$. It is convenient
     to work in the interaction picture.\footnote{The special role of $H_0$ in the time dependence of states and operators in the interaction picture is analogous to the choice of action-angle variables for the free rather than the full Hamiltonian in the version of canonical perturbation theory used by Van Vleck. This is what lies behind the close similarities between the calculations in this section and those in secs.\ 5.1 and 6.2.} The state $|r,t>_{\rm int}$ in the interaction picture is related to the state $|r,t>$ in the Schr\"{o}dinger picture via:
\begin{equation}
|r,t>_{\rm int} \equiv e^{iH_0t/\hbar} |r,t>.
\label{5.1}
\end{equation}
   An operator $O_{\rm int}(t)$ in the interaction picture is related to the corresponding operator $O$ in the Schr\"{o}dinger picture via 
     \begin{equation}
     \label{eq:timedepint}
       O_{\rm int}(t) \equiv e^{iH_0 t/\hbar}Oe^{-iH_0 t/\hbar}.
       \label{5.2}
       \end{equation}
It follows that expectation values are the same in the two pictures:
\begin{equation}
_{\rm int}\!\!<r,t|O_{\rm int}(t) |r,t>_{\rm int} = <r,t|O|r,t>.
\end{equation} 
The evolution of the states in the interaction picture is given by:
\begin{eqnarray}
\frac{\partial}{\partial t} |r,t>_{\rm int} & = & \frac{i}{\hbar} e^{iH_0t/\hbar} H_0 |r,t> + e^{iH_0t/\hbar} 
\frac{\partial}{\partial t} |r,t> \nonumber \\ & = & \frac{i}{\hbar}
e^{iH_0t/\hbar} \left( H_0 - H \right) |r,t>, \label{evol}
\end{eqnarray}
where in the last step, we used the Schr\"{o}dinger equation
\begin{equation}
\frac{\partial}{\partial t} |r,t> = - \frac{iH}{\hbar} |r,t>. 
\end{equation}
Since $H_0 - H = -V(t)$ (see eq.\ (\ref{eq:quantHam})), we can write eq.\ (\ref{evol}) as:
       \begin{eqnarray}
      \frac{\partial}{\partial t}|r,t>_{\rm int} & = & - \frac{i}{\hbar} e^{iH_0t/\hbar} V(t)  e^{-iH_0t/\hbar} |r,t>_{\rm int}
       \nonumber \\ & = & -\frac{i}{\hbar} V_{\rm int}(t)|r,t>_{\rm int}, \label{eq:intstate}
       \end{eqnarray}
where we used eqs.\ (\ref{5.1})--(\ref{5.2}).
    To first order in $V_{\rm int}(t)$ (i.e., to first order in the field $E$), the solution of  (\ref{eq:intstate}) is
    \begin{eqnarray}
    \label{eq:intstate1}
     |r,t>_{\rm int} &=& |r,0>_{\rm int} -\frac{i}{\hbar}\int_{0}^{t} d\tau V_{\rm int}(\tau) |r,0>_{\rm int}  \nonumber \\
     \label{eq:intstate2}
         &=& |r,0>_{\rm int} -\frac{ieE}{\hbar}\int_{0}^{t} d\tau x_{\rm int}(\tau)\cos{\omega \tau}|r,0>_{\rm int}.
         \end{eqnarray}
  At $t=0$ the states (and operators) in the interaction picture coincide with those in the Schr\"{o}dinger picture. From now on  we thus simply write $|r>$ for $|r,0>_{\rm int}$. The dual (`bra') of the vector (`ket') in eq.\ (\ref{eq:intstate2}) is:
  \begin{equation}
_{\rm int}\!\!<r,t| =  <r| + \frac{ieE}{\hbar}\int_{0}^{t} d\tau \cos{\omega \tau}<r| x_{\rm int}(\tau).
\label{bra}
  \end{equation}
  To find the dipole moment $P_r(t)$  of the system in state $r$ to first order in $E$, we calculate the first-order contribution to the expectation value of the displacement $<\Delta x>_r$ in the state $r$ induced by the field $E$:
  \begin{equation}
 <\Delta x>_r \equiv \; _{\rm int}\!\!<r,t|x_{\rm int}(t)|r,t>_{\rm int} - <r|x_{\rm int}(t)|r>.
  \end{equation}    
 Inserting eqs.\ (\ref{eq:intstate2})--(\ref{bra}) into this expression, we find:
  \begin{equation}
  \label{eq:qmpolar1}
   <\Delta x>_r = \frac{ieE}{\hbar}  \int_{0}^{t} d\tau <r|
   \left\{ x_{\rm int}(\tau)x_{\rm int}(t) - x_{\rm int}(t) x_{\rm int}(\tau) \right\} |r>\cos{\omega\tau}.
    \end{equation}
 Writing $\cos{\omega\tau}=\frac{1}{2}(e^{i\omega\tau}+e^{-i\omega\tau})$, and inserting
 a complete set of eigenstates of the unperturbed Hamiltonian  $H_0$ ($1=\sum_{s}|s><s|$) between the two coordinate
 operators, we obtain
 \begin{eqnarray}
  <\Delta x>_r &=& \frac{ieE}{2\hbar} \sum_{s}\int_{0}^{t} d\tau \left( <r|e^{iH_{0}\tau/\hbar}xe^{-iH_{0}\tau/\hbar}
  |s><s|e^{iH_{0}t/\hbar}xe^{-iH_{0}t/\hbar}|r> \right. \nonumber  \\
  \label{eq:qmpolar2}
  & & \;\;\;\;\;\;  \left. - <r|e^{iH_{0}t/\hbar}xe^{-iH_{0}t/\hbar}|s><s|e^{iH_{0}\tau/\hbar}xe^{-iH_{0}\tau/\hbar}|r> \right)
  e^{i\omega\tau} \nonumber  \\ & & \;\;\;\;\;\; + \;\;\; (\omega\rightarrow -\omega) \nonumber \\
  &=& \frac{ieE}{2\hbar} \sum_{s}\int_{0}^{t} d\tau \left( e^{i(E_r-E_s+\hbar\omega)\tau/\hbar}e^{i(E_s-E_r)t/\hbar}
  \right. \nonumber \\
  \label{eq:qmpolar3}
  && \;\;\;\;\;\; \left. - e^{i(E_r-E_s)t/\hbar}e^{i(E_s-E_r+\hbar\omega)\tau/\hbar} \right) <r|x|s> <s|x|r> \\ & &
 \;\;\;\;\;\; + \;\;\; (\omega\rightarrow -\omega). \nonumber
  \end{eqnarray}
We introduce the notation  $X_{rs}\equiv <r|x|s>$ for the matrix elements of the coordinate operator. Note that these matrix elements  in eq.\ (\ref{eq:qmpolar3}) are accompanied by time-development phases $e^{i(E_r-E_s)t/\hbar}$ of {\it purely
 harmonic form}: they are the precise correlates in modern quantum mechanics of the substitute oscillators of \citet{Ladenburg and Reiche 1923} or, equivalently, the virtual oscillators of BKS, as was clearly realized, for instance, by \citet{Lande 1926} (see also the discussion at the end of sec.\ 4.3).
 
 Performing the time integral in eq.\ (\ref{eq:qmpolar3}), we find
 \begin{eqnarray}
  <\Delta x>_r &=& \frac{eE}{2} \sum_{s} \left[ 
    \frac{e^{i(E_r - E_s + \hbar \omega)t/\hbar} - 1}{E_r - E_s + \hbar \omega} e^{i(E_s - E_r)t/\hbar}
   \right. \nonumber \\ & & \;\;\;\;\;\;\;\;\;\;\;\;\;\;\;\;\;\;\;\;\;
  \left.
  - \frac{e^{i(E_s - E_r + \hbar \omega)t/\hbar} - 1}{E_s - E_r + \hbar \omega} e^{i(E_r - E_s)t/\hbar}
   \right] X_{rs}X_{sr}  \\ & & + \;\;\; (\omega\rightarrow -\omega). \nonumber
 \end{eqnarray}
 (cf.\ eqs.\ (\ref{eq:2.25}) and (\ref{eq:2.27}) in sec.\ 6.1 and eqs.\ (\ref{eq:3.23}) and (\ref{eq:3.27}) in sec.\ 5.2). 
The coherent terms in $<\Delta x>_r$, i.e. the terms with a time-dependence $e^{\pm i\omega t}$ (cf.\ eq.\ (\ref{eq:2.30}) in sec.\ 6.1 and eq.\ (\ref{eq:3.30}) in sec.\ 5.2), are:
 \begin{eqnarray}
 \label{eq:qmpolar4}
 <\Delta x_{\rm coh}>_r  & = & \frac{eE}{2}  \sum_{s} X_{rs}X_{sr}  e^{i\omega t} \left[ \frac{1}{E_r -E_s +\hbar\omega}  - \frac{1}{E_s -E_r +\hbar\omega} \right] \label{x_coh} \\ & & \;\;\; + \;\;\; (\omega\rightarrow -\omega). \nonumber
 \end{eqnarray}
 Using the Bohr frequency condition $\hbar \omega_{rs} = E_r - E_s$, we can write the expression in square brackets in eq.\ (\ref{x_coh}) as:
 \begin{equation}
 \frac{1}{\hbar \omega_{rs} +\hbar\omega} -
 \frac{1}{\hbar \omega_{sr} +\hbar\omega} = \frac{2 \omega_{rs}}{\hbar (\omega^2_{rs} - \omega^2)}.
 \end{equation}
 Inserting this result into eq.\ (\ref{x_coh}) and noting that the terms proportional to $\sin{\omega t}$ vanish, we find the following result for the dipole moment of the system in state $r$ (cf.\ eq.\ (6) or eq.\ (\ref{eq:2.32a}))
\begin{equation}
P_r(t) = -e <\Delta x_{\rm coh}>_r =  \frac{2e^2E}{\hbar}\sum_{s}  \frac{\omega_{sr}X_{rs}X_{sr}}{\omega^2_{sr} - \omega^2} 
 \cos{\omega t}.
\end{equation}
The sum over $s$ can  naturally be separated into states $s$ of higher energy than $r$, with $\omega_{sr}>0$, and states
  $t$ of lower energy, with $\omega_{rt}>0$ ($\omega_{rt}=0$ for $r=t$):
  \begin{equation}
  \label{eq:qmpolarfinal}
   P_r=\frac{2e^{2}E}{\hbar} \left( \sum_{s}\frac{\omega_{sr}X_{sr}X_{rs}}{\omega_{sr}^{2}-\omega^{2}}
   -\sum_{t}\frac{\omega_{rt}X_{rt}X_{tr}}{\omega_{rt}^{2}-\omega^{2}} \right) \cos{\omega t}.
   \end{equation}
   If we recall the correspondence principle for emission (\ref{eq:3.15}), and identify
   $D_{s}^{2}$ with $3(X^{s}_{\tau})^{2}=12A_{\tau}A_{-\tau}$ and the Fourier coefficients
   $A_{\tau}\rightarrow X_{sr}$, $A_{-\tau}\rightarrow X_{rs}$ we get
   \begin{equation}
     A_{s\rightarrow r}=\frac{64\pi^{4}e^{2}}{hc^{3}}\nu_{sr}^{3}X_{sr}X_{rs}, \label{A_sr}
     \end{equation}
     whence we recover the original form (\ref{eq:2.43}) of the dispersion formula
     \begin{equation}
     \label{eq:Kramfinal}
     P_r=\frac{c^{3}}{32\pi^{4}}E\cos{\omega t} \left( \sum_{s}\frac{A_{s\rightarrow r}}{\nu_{sr}^{2}(\nu_{sr}^{2}-\nu^{2})}-\sum_{t}\frac{A_{r\rightarrow t}}{\nu_{rt}^{2}(\nu_{rt}^{2}-\nu^{2})} \right).
     \end{equation}
  Of course, the above identification of classical Fourier components with matrix elements
  of the position operator is at the core of Heisenberg's 1925 breakthrough.

    Returning for a moment to eq.\ (\ref{eq:qmpolarfinal}), we see that in the Thomson limit where the
    frequency of incident radiation far exceeds the difference frequencies $\omega_{rs}$ for
    the electron states $r,s$,\footnote{Or, alternatively, when the incident photon energy far exceeds the
    energy needed to ionize the electron, so that the latter can be regarded as essentially a
    free, unbound particle.} the polarization $P_r$  becomes asymptotically
    \begin{equation}
    \label{eq:ThomKuhnlim1}
      P_{r} \simeq -\frac{2e^{2}E}{\hbar\omega^{2}}(\sum_{s}\omega_{sr}X_{sr}X_{rs}-\sum_{t}\omega_{rt}X_{rt}X_{tr} )\cos{\omega t}.
      \end{equation}
      The preceding equation is in content identical with the next to last (unnumbered)  equation   in                     
      sec.\ 2 in \citep{Heisenberg 1925c}, 
      where the Kramers dispersion theory is explicitly invoked. For large frequencies, we expect the polarization to approach our previously derived result (see eq.\ (6) or eq.\ (\ref{eq:2.32a})) for the polarization of a charged harmonic oscillator in the limit where $\nu >> \nu_0$:\footnote{This result is obtained in \citep{Kuhn 1925} by equating the energy scattered by an electron in the Thomson limit to the radiation emitted by an oscillating dipole according to the Larmor formula.}
        \begin{equation}
        \label{eq:ThomPr}
          P_{r} = -\frac{e^{2}E}{m\omega^{2}}\cos{\omega t},
          \end{equation}
Comparing eq.\ (\ref{eq:ThomKuhnlim1}) with eq.\ (\ref{eq:ThomPr}) we find eq.\ (16) in \citep{Heisenberg 1925c}:
          \begin{equation}
          \label{eq:Heis16}
          h=4\pi m(\sum_{s}\omega_{sr}X_{sr}X_{rs}-\sum_{t}\omega_{rt}X_{rt}X_{tr}).
          \end{equation}     
          This result is first obtained by Heisenberg from the Bohr-Sommerfeld quantization
          condition by applying the quantum-theoretical transcription procedure, which was 
          introduced in sec.\ 1 of the {\it Umdeutung} paper and had been inspired by dispersion theory. It replaces the Bohr-Sommerfeld condition as the fundamental quantization constraint in Heisenberg's new theory. That the same result
          can be obtained directly from the high-frequency limit of the Kramers dispersion
          formula is clearly regarded by Heisenberg as strong evidence for the validity
          of his transcription procedure. 
      Using eq.\ (\ref{eq:Heis16}), together with the formal
      transcription of the classical equation of motion, $\ddot{x}+f(x)=0$  (eq.\ (11) of the {\it Umdeutung}
      paper), 
      \citet{Heisenberg 1925c} asserts the possibility
      of ``a complete determination not only of frequencies and energy values, but also of
      quantum-theoretical transition probabilities" (p.\ 268).  As Heisenberg points out, eq.\ (\ref{eq:Heis16})
      is completely equivalent to the sum rules for oscillator strengths derived by
      \citet{Thomas 1925} and \citet{Kuhn 1925}.\footnote{Heisenberg's logic is slightly different from ours. Instead of pointing out that the high-frequency limit (\ref{eq:ThomKuhnlim1}) of the Kramers dispersion formula and the well-established classical result (\ref{eq:ThomPr}) imply Heisenberg's quantization condition (\ref{eq:Heis16}), \citet[pp.\ 269--270]{Heisenberg 1925c} points out that eqs.\ (\ref{eq:Heis16}) and (\ref{eq:ThomKuhnlim1}) imply eq.\ (\ref{eq:ThomPr}). This is only a cosmetic difference. The point of the exercise is still to show that the new quantization condition, found through {\it Umdeutung} of the derivative of the Bohr-Sommerfeld condition, follows from well-established results in Kramers' dispersion theory and classical electrodynamics. We are nonetheless grateful  to Christoph Lehner for alerting us to this point.}
      
  The realization that eq.\ (\ref{eq:Heis16}) is equivalent to (the diagonal matrix elements of) the fundamental 
   commutator relation $[P,X]= \hbar/i$ of modern quantum theory came shortly
   after this, in the work of \citet{Born and Jordan 1925b}. The recognition of    eq.\ (\ref{eq:Heis16})
   as a commutator is mathematically obscured by the separation of the sum into
   states  higher ($s$) and lower ($t$) than the given state $r$---a separation
   which is very natural given the history of the Kramers dispersion
   formula.  If Heisenberg had
   applied his own transcription rules for associating classical variables with 
   quantum two-index quantities to the momentum  $P\equiv m\dot{X}$ in the unnumbered
   equation immediately following (13) in the {\it Umdeutung} paper \citep[p.\ 267]{Heisenberg 1925c}, he would have found (using modern matrix notation):\footnote{Following Heisenberg's procedure in the {\it Umdeutung} paper for translating classical equations into quantum-mechanical ones, we would translate his  classical equation for momentum, $m\dot{x} = m \sum_\alpha a_\alpha(n) i \alpha \omega_n e^{i\alpha \omega_n t}$, into the following quantum-mechanical equation: $P(n, n + \alpha) = i m a(n, n + \alpha) \omega(n, n + \alpha)$. In modern notation, this becomes: $P_{rs} = i m X_{rs}  \omega_{rs}$ (no summation).} 
   \begin{equation}
   \label{eq:momop}
     P_{rs} = im\omega_{rs}X_{rs}.
     \end{equation}
     That Heisenberg did not write down this equation is, as we suggested above, because he was thinking in terms of the Lagrange rather than the Hamilton formalism.
     Rewriting eq.\ (\ref{eq:Heis16}) as a {\it single} sum over all states $s$, but splitting the sum into
     two equal pieces via the identity $2\omega_{sr}=\omega_{sr}-\omega_{rs}$, we find
     \begin{eqnarray}
     h&=&4\pi m\sum_{s}\omega_{sr}X_{rs}X_{sr} \nonumber \\
      &=& 2\pi m\sum_{s}(X_{rs}\omega_{sr}X_{sr}-\omega_{rs}X_{rs}X_{sr}) \label{eq:pqcomm} \\
      &=& -2\pi i \sum_{s}(X_{rs}P_{sr}-P_{rs}X_{sr}), \nonumber
      \end{eqnarray}
     where in the last step we used eq.\ (\ref{eq:momop}). In modern notation, this last expression is immediately recognized as the diagonal matrix element
      of the fundamental commutator $[X,P]= i\hbar$:
      \begin{eqnarray}
      \label{eq:pqcommdiag}
      i\frac{h}{2\pi} &= &<r|XP-PX|r>   \nonumber     \\
      &= & \sum_{s}(<r|X|s><s|P|r>-<r|P|s><s|P|r>).
      \end{eqnarray}
      Although Heisenberg recognized the significance of the noncommutativity of 
      quantum-theoretic quantities in his formalism (see the last three paragraphs of
      sec.\ 1), the simplicity of $x(t)p(t)-p(t)x(t)$ implied by his fundamental
      quantization relation \ (\ref{eq:Heis16}) eluded him. He was thinking in terms of velocity rather than momentum. Moreover, even if he had been thinking in terms of momentum, the origin of his quantization condition in dispersion theory might well have prevented him from rewriting the summations the way we did in  eq.\ (\ref{eq:pqcomm}).
      
\subsection{Spontaneous emission}

To begin with, we note that we are dealing throughout with the dipole approximation,
which is implicit in the 1924 work, corresponding to the regime where the wavelength
of light is much larger than atomic dimensions (or equivalently, where photon momentum is much smaller than electron
momentum). Once again, note that the notation of \citep[eq.\ (1)]{Van Vleck 1924b},
\begin{eqnarray}
 x & = & \sum_{\tau_{1}\tau_{2}\tau_{3}}X(\tau_{1},\tau_{2},\tau_{3})\cos{ \{ 2\pi(\tau_{1}\omega_{1}
+\tau_{2}\omega_{2}+\tau_{3}\omega_{3})t+ \ldots \} } \nonumber \\
 & = & \sum \left\{ \frac{1}{2}X(\tau_{1},\tau_{2},\tau_{3})e^{+2\pi i(\tau_{1}\omega_{1}+\tau_{2}\omega_{2}+\tau_{3}\omega_{3})t+ \ldots}  \right.  \\ & & \;\;\;\;\;\;\;\;\;\;\;\;\;\;\;\;\;\;\;\;\;\;\;\; \left.
           +\frac{1}{2}X(\tau_{1},\tau_{2},\tau_{3})e^{-2\pi i(\tau_{1}\omega_{1}+\tau_{2}\omega_{2}+\tau_{3}\omega_{3})t+ \ldots} \right\}, \nonumber
\end{eqnarray}
implies that van Vleck's $D^{2}=X^2+Y^2+Z^2$ \citep[line following eq.\ (8)]{Van Vleck 1924b}
corresponds to four times the square of the matrix element of the quantum position operator
appearing in the dipole transition formulas of modern quantum mechanics. For the latter we shall follow
the treatment of \citep[Ch.\ 13]{Baym 1969}.

 In the dipole approximation, the spontaneously emitted power per unit solid angle is given
by \citep[p.\ 282, eq.\ 13--100]{Baym 1969}, for emitted light of polarization vector $\vec{\lambda}$, in a transition
from state $r$ to state $s$:
\begin{eqnarray}
\label{eq:baymsp}
  \frac{dP}{d\Omega}&=& \frac{\omega^{4}e^{2}}{2\pi c^{3}}<r|\vec{\lambda}\cdot \vec{x}|s><s|\vec{\lambda}\cdot\vec{x}|r>  \nonumber \\
  &=& \sum_{i,j=1}^{3} \frac{\omega^{4}e^2}{2\pi c^{3}}\lambda_{i}\lambda_{j}<r|x_{i}|s><s|x_{j}|r>.
\end{eqnarray}
Here (unlike Baym) we take real polarization vectors $\vec{\lambda}$ (plane polarized)
rather than complex (circularly polarized) ones as our basis. We want the total spontaneously
emitted power in any event, summed over the two possible polarizations for any momentum
vector $\vec{k}$ of the emitted photon (so the basis of photon states is irrelevant).
This requires the polarization sum
\begin{equation}
\label{eq:polsum}
  \sum_{\lambda=1}^{2} \lambda_{i}\lambda_{j} = \delta_{ij}-\hat{k}_{i}\hat{k}_{j},\;\;\;(i,j=1,2,3),
\end{equation}
which follows from the fact that the two polarization vectors are any pair of orthogonal unit vectors perpendicular
to the unit vector $\hat{k}$ along the photon direction. Finally, we want the total power
emitted in any direction, so the polarization sum (\ref{eq:polsum}) must be integrated over
all solid angles:
\begin{equation}
\label{eq:intpolsum}
  \int d\Omega_{\hat{k}}(\delta_{ij}-\hat{k}_{i}\hat{k}_{j})= 4\pi(\frac{2}{3}\delta_{ij}).
\end{equation}
The Einstein coefficient $A_{r\rightarrow s}$ in \citep[eqs.\ (5) and (9)]{Van Vleck 1924b} refers to a rate of
photon emission (not energy emission) so we must divide eq.\ (\ref{eq:baymsp}) by $\hbar\omega$.
Putting together the above results (and switching to $\nu=\omega/2\pi$), we find:
\begin{equation}
A_{r\rightarrow s} = \frac{1}{\hbar\omega}\int d\Omega_{\hat{k}} \frac{dP}{d\Omega_{\hat{k}}} =
\frac{\omega^{4}e^2}{2\pi \hbar\omega c^{3}}\frac{8\pi}{3}\sum_{i}<r|x_{i}|s><s|x_{i}|r>.
\end{equation}
Using the notation $X_{rs} \equiv <r|x|s>$, etc. for the matrix elements of position introduced above we can rewrite this as:
\begin{equation}
A_{r\rightarrow s} = \frac{\omega^4 e^2}{2 \pi \hbar \omega c^3} \frac{8\pi}{3} \left( |X_{rs}|^2 + |Y_{rs}|^2 + |Z_{rs}|^2 \right).
\end{equation}
Replacing the matrix elements $X_{rs}$, $Y_{rs}$, and $Z_{rs}$ by the amplitude $D_r$ in the correspondence limit as indicated in the preceding section (cf.\ the remarks preceding eq.\ (\ref{A_sr})) and substituting $\omega = 2\pi \nu$, we arrive at:
\begin{equation}
A_{r\rightarrow s} = \frac{16\pi^{4}e^{2}\nu^{3}}{3hc^{3}}D_{r}^{2}.  \label{emission}
\end{equation}
$D_{r}^2$ is the amplitude defined by \citep{Van Vleck 1924b} immediately following eq.\ (8), to be replaced by  $D_{r}(\tau_{1},\tau_{2},\tau_{3})^{2}$ in eq.\ (9), with which eq.\ (\ref{emission}) is
seen to be identical.

\subsection{Absorption}

 The Einstein formula for absorption \citep[eq.\ (6)]{Van Vleck 1924b}, when combined with the
stimulated emission (``negative absorption") term to yield (ibid., eq.\ 15)), leads directly to
the correspondence limit result (ibid., eq.\ (16)). Here, we check the identity of eq.\ (15) in
\citep{Van Vleck 1924b}  (more
precisely, the unnumbered  equation immediately following this one)   with 
the modern absorption calculation given in \citep{Baym 1969}.
  For the rate of absorption of light leading to a transition from state $s$ to (higher) state $r$,
\citep[eq.\ 13--40]{Baym 1969} reads (in dipole approximation, $\vec{j}_{\vec{k}}\rightarrow \vec{p}/m$):
\begin{equation}
\label{eq:baymabs}
 \Gamma^{\rm abs}_{s\rightarrow r}=\frac{2\pi e^2}{\hbar^{2}c^2}
\frac{\omega^{2}}{(2\pi c)^{3}}\int d\Omega_{\hat{k}}\sum_{\lambda}<s|\vec{\lambda}\cdot\frac{\vec{p}}{m}|r><r|\vec{\lambda}\cdot\frac{\vec{p}}{m}|s>|A_{\vec{k}\vec{\lambda}}|^{2}.
\end{equation}
As usual, in dipole approximation we can use \citep[eq.\ 13--98]{Baym 1969} to replace matrix
elements of the momentum operator with those of the coordinate operator (using the equations
of motion). For Hamiltonians of the form $H=(\vec{p}^{2}/2m)+V(\vec{x})$,
\begin{equation}
  [H,x_{j}] = \frac{1}{2m}[p_{i}p_{i},x_{j}] =  \frac{1}{m}p_{i}[p_{i},x_{j}] 
      = \frac{p_{i}}{m}\frac{\hbar}{i}\delta_{ij} = \frac{\hbar}{i}\frac{p_{j}}{m},
\end{equation}
     whence
\begin{eqnarray}
\label{eq:XtoP}
<r|\frac{\vec{p}}{m}|s>&=&\frac{i}{\hbar}<r|[H,\vec{x}]|s> \nonumber \\
&=&\frac{i}{\hbar}(E_{r}-E_{s})<r|\vec{x}|s> \label{matrix-element} \\
&=&i\omega<r|\vec{x}|s>, \nonumber
\end{eqnarray} 
where $\hbar\omega=E_{r}-E_{s}$. Once again, in eq.\ (\ref{matrix-element}), we see the ``monstrous"
difference frequencies characteristic of quantum theory, which wreaked havoc on
classical interpretations of radiation phenomena, making their appearance in the modern formalism.
 Accordingly,  eq.\ (\ref{eq:baymabs}) becomes
\begin{equation}
\label{eq:baymabs2}
 \Gamma^{\rm abs}_{s\rightarrow r}=\frac{2\pi e^2}{\hbar^{2}c^2}
\frac{\omega^{4}}{(2\pi c)^{3}}\int d\Omega_{\hat{k}}\sum_{\lambda}<s|\lambda_{i}x_{i}|r><r|\lambda_{j}x_{j}|s>|A_{\vec{k}\vec{\lambda}}|^{2}.
\end{equation}
Now we are going to assume that the ambient light is unpolarized and isotropic so that the squared
amplitude $|A_{\vec{k}\vec{\lambda}}|^2$ is in fact independent of $\lambda,\hat{k}$, and the
only angular dependence comes in via the polarization vectors. The angle average
of the polarization sum in eq.\ (\ref{eq:baymabs2}) can then be performed as in eq.\ (\ref{eq:intpolsum})
to yield
\begin{equation}
\label{eq:baymabs3}
 \Gamma^{\rm abs}_{s\rightarrow r}=\frac{4\pi e^2}{3\hbar^{2}c^2}
\frac{\omega^{4}}{(2\pi c)^{3}}<s|x_{i}|r><r|x_{i}|s>\int d\Omega_{\hat{k}}|A_{\vec{k}\vec{\lambda}}|^{2}.
\end{equation}
 Next, we need to establish the relation between the squared mode amplitudes $|A_{\vec{k}\vec{\lambda}}|^{2}$ and the specific energy density function $\rho(\nu)$ defined as the energy per unit
volume per unit frequency interval. The mode amplitudes $A_{\vec{k}\vec{\lambda}}$ correspond
to discrete modes for electromagnetic radiation in a box of volume $V$, with each mode contributing
energy density 
\begin{equation}
\frac{1}{V}|A_{\vec{k}\vec{\lambda}}|^{2}\frac{\omega}{2\pi c^{2}}
\end{equation} 
\citep[eq.\ 13--14]{Baym 1969}. As the box volume goes to infinity we have the usual correspondence
\begin{equation}
 \frac{1}{V}\sum_{k} \rightarrow \int \frac{k^{2}dkd\Omega_{\hat{k}}}{(2\pi)^{3}},
\end{equation}
so that the total energy density {\it between frequency $\nu$ and frequency $\nu+\Delta\nu$}
is
\begin{eqnarray}
  \rho(\nu)\Delta\nu &=& \frac{1}{V}\sum_{2\pi\nu<kc<2\pi(\nu+\Delta\nu)}2|A_{\vec{k}\vec{\lambda}}|^{2}\frac{\omega^{2}}{2\pi c^{2}} \nonumber \\
\label{eq:rho1}
  &\rightarrow& \frac{1}{(2\pi)^{3}}\int d\Omega_{\hat{k}}\int_{2\pi\nu/c}^{2\pi(\nu+\Delta\nu)/c}dk\;\;
 k^{2}\frac{\omega^{2}}{2\pi c^{2}}2|A_{\vec{k}\vec{\lambda}}|^{2}.
\end{eqnarray}
Note that although we continue to write the mode amplitudes $A_{\vec{k}\vec{\lambda}}$ as 
depending on polarization and momentum vector of the photon, we are really assuming that
there is no dependence on the polarization or photon {\it direction}. Hence the factor
of 2, with no remaining sum over $\lambda$. Eq. (\ref{eq:rho1})
gives
\begin{equation}
 \rho(\nu)\Delta\nu = \frac{1}{(2\pi)^{3}}\frac{2\pi}{c}k^{2}\frac{\omega^{2}}{2\pi c^{2}}2
 \int d\Omega_{\hat{k}}|A_{\vec{k}\vec{\lambda}}|^{2} \Delta\nu,
\end{equation}
or, equivalently
\begin{equation}
\label{eq:asq}
 \int d\Omega_{\hat{k}}|A_{\vec{k}\vec{\lambda}}|^{2} = \frac{4\pi^{3}c^{5}}{\omega^{4}}\rho(\nu).
\end{equation}
Inserting eq.\ (\ref{eq:asq}) into eq.\ (\ref{eq:baymabs3}) and multiplying by $\hbar\omega$ to get
the rate of energy absorption (instead of the number rate of photon absorption)  we find,
using the usual association of squares of matrix elements of the position operator to the
classical orbit amplitude $\frac{1}{4}D_r^2$,
\begin{eqnarray}
 \hbar\omega\Gamma^{\rm abs}_{s\rightarrow r} &=& \frac{4\pi e^{2}\omega}{3\hbar c^2}\frac{\omega^{4}}
{(2\pi c)^{3}}\frac{4\pi^{3}c^{5}}{\omega^{4}}\rho(\nu)\frac{1}{4}D_{r}^{2} \nonumber \\
  &=& \frac{2\pi^{3}e^{2}}{3h}\nu\rho(\nu)D_{r}^{2},
\end{eqnarray}
which coincides with the first term in van Vleck's equation \citep[the equation following eq.\ (15)]{Van Vleck 1924b} for the part
of the total absorption rate due to upward transitions. Of course, the second  (negative
absorption, or stimulated emission) term is
of exactly the same form (with a minus sign) due to the symmetry of the Einstein $B$ coefficients.

\section{Conclusion}

Our study of Van Vleck's two-part paper on the application of the correspondence principle to the interaction of matter and radiation \citep{Van Vleck 1924b, Van Vleck 1924c} has led us to consider three clusters of questions. First, there are questions about the paper itself. What made Van Vleck decide to work in this area? He had not published on radiation theory before. 
And---as one is inevitably tempted to ask---why did Van Vleck not take the next step and arrive at something like matrix mechanics? That gets us to the second cluster of questions, about the developments in quantum theory that provide the natural context for Van Vleck's work, especially the transition of the old quantum theory of Bohr and Sommerfeld to matrix mechanics. What was important for this development and what was not? The third group of questions concerns the relative importance of American contributions to these developments. In this final section we collect the (partial) answers we have found to these biographical, conceptual, and sociological questions. 

Let us first dispose of the issue of American contributions to early quantum theory. Since we focused on the work of only two individuals, Van Vleck and Slater, we are in no position to draw strong conclusions. Still, it seems safe to say that our study supports the thesis of Sam \citet{Schweber 1986} and others that, by the early 1920s, the United States had a homegrown tradition in quantum theory, which, to be sure, was reinforced, but certainly not created by the influx of European \'{e}migr\'{e}s in the 1930s. We are less sanguine about the thesis of Alexi \citet{Assmus 1992b} that American theorists contributed mainly to molecular rather than to atomic physics, although she may be right that Slater and Van Vleck are just the exception to the rule (see sec.\ 2.4). However, we did come across several other contributions (some admittedly minor) to atomic theory by Americans (Breit, Davisson, Hoyt, Kemble) or by Europeans working in America (Epstein, Swann). And we do want to emphasize that the contributions to atomic theory by our main protagonists were absolutely first rate, even if they did not always receive the recognition they deserved from their European colleagues (see the correspondence between Born and Van Vleck cited in secs.\ 2.4 and 5.2). The quickly refuted but highly influential Bohr-Kramers-Slater (BKS) theory was built around Slater's idea of a virtual radiation field emitted by an atom while in a stationary state (see sec.\ 4.1).  The derivation of a correspondence principle of absorption for a general non-degenerate multiply-periodic system, the centerpiece of \citep{Van Vleck 1924b, Van Vleck 1924c},  is a {\it tour de force} that may well have been the most sophisticated application of the correspondence principle in the old quantum theory. All in all, the Americans had definitely established a presence in atomic theory by the early 1920s.  In the period we examined, they were certainly more prominent than the British, not to mention the French. Ultimately, however, the decisive steps were taken in Europe, not in the United States.

This brings us to the question of why Van Vleck stopped short of these decisive steps. Before we offer our best guess as to why Van Vleck did not do what he did not do, we want to say a few words about why he did what he did. His papers on the correspondence principle for absorption \citep{Van Vleck 1924a, Van Vleck 1924b, Van Vleck 1924c} constitute his first foray into quantum radiation theory. His earlier publications had dealt with such topics as the extension of Bohr's model of the atom to helium and the specific heat of molecular hydrogen.  The formulation of a correspondence principle for absorption, Van Vleck told Kuhn in his interview for the AHQP in 1963, had been triggered by a comment of his Minnesota colleague Breit (see also Van Vleck, 1924a, p.\ 28). Breit's remark, we conjectured (in sec.\ 5.3), may have directed Van Vleck to the work of \citet{Ladenburg and Reiche 1923}, who proposed quantum formulae for emission, absorption, and dispersion, invoking but not always correctly implementing the correspondence principle. Van Vleck constructed his own quantum formulae for emission and absorption and used his considerable expertise in classical mechanics to show that these formulae as well as the Kramers dispersion formula merged with the classical formulae in the limit of high quantum numbers. 

So why did Van Vleck not take the next step? The trivial explanation is that he was too busy working on his {\it Bulletin} for the {\it National Research Council} on the old quantum theory \citep{Van Vleck 1926a} to pursue his own research. But even if he had not been burdened by this {\it Bulletin}, we seriously doubt that Van Vleck would have done what Heisenberg did---as he himself acknowledged both in a biographical statement prepared for the AHQP and in his interview for the project (see sec.\  1.2). Van Vleck, it seems, was too wedded to the orbits of the Bohr-Sommerfeld theory to completely discard them, a prerequisite for Heisenberg's {\it Umdeutung}. This is clear at several points in \citep{Van Vleck 1924b}. At the end of sec.\ 1, for instance, we find a formula expressing the Einstein coefficient $A_{r \rightarrow s}$ as an average over the frequencies of orbits, not allowed by the Bohr-Sommerfeld quantization condition, between the initial state $r$ and the final state $s$. Sec.\ 2 of the paper is devoted to ``a correspondence principle for {\it orbital distortions}" \citep[p.\ 334, our emphasis]{Van Vleck 1924b}. On the issue of how seriously one should take the orbits of the Bohr-Sommerfeld theory, Van Vleck might have benefited from direct contact with the Europeans. He had the distinct disadvantage of {\it reading} Sommerfeld instead of {\it talking} to Bohr and his circle.\footnote{According to Alexi \citet[pp.\ 8, 15]{Assmus 1992b}, Americans had a tendency to follow Sommerfeld rather than Bohr anyway.} Bohr and Pauli certainly prepared Heisenberg for the step of leaving orbits behind.

The emphasis on observable quantities in the {\it Umdeutung} paper, however, struck a chord with Van Vleck, who had been primed for such a positivist turn by his Harvard teacher Bridgman.\footnote{In the biographical note written for the AHQP, Van Vleck wrote: ``I suspect that Bridgman's operational philosophy may have subconsciously influenced my approach to theoretical physics." At a ceremony honoring Bridgman's 1946 Nobel prize, Slater went as far as suggesting a genetic  link between Bridgman's operationalism and Heisenberg's uncertainty principle! \citet{Schweber 1990} quotes Slater as saying on this occasion: 
``It is very likely that this principle, so much like Bridgman's attitude, is actually derived to a very considerable extent from Bridgman's thinking" (p.\ 391).} Explaining the new quantum mechanics in {\it Chemical Reviews} in 1928,\footnote{For the benefit of the chemists, \citet{Van Vleck 1928a} compared a matrix to a baseball schedule: ``the entry in row 3 and column 2, for instance, gives information about a transition between a 3 and 2 quantum state, just as the analogous baseball entry does about the meetings between teams 3 and 2" (p.\ 469).} he wrote:
\begin{quotation}
Heisenberg's epoch-making development of the matrix theory was spurred by Born's repeated emphasis to his colleagues at G\"ottingen that the reason the old quantum theory was then (1925) failing was that we were all too anxious to use the same concepts of space and time within the atom as in ordinary measurable large-scale events. 
\ldots the concepts of distance and time have a meaning only when we tell how they can be measured. This is very nicely emphasized in Bridgman's recent book, ``The Logic of Modern Physics" [Bridgman, 1927] \ldots one cannot use a meter stick to measure the diameter of an atom, or an alarm clock to record when an electron is at the perihelion of its orbit. Consequently we must not be surprised if within the atom the correlation of space and time is something which cannot be visualized, and that models cannot be constructed with the same kind of mechanics as Henry Ford uses in designing an automobile. \ldots The goal of theoretical physics and chemistry must ever be to explain observable rather than unobservable phenomena \ldots What the physicist observes about an atom is primarily its radiations \ldots We may say that we have a sound atomic theory when we have a set of a small number of mathematical postulates from which these observed things can be calculated correctly, even though it forces us to discard the usual space-time models \citep[p.\ 468]{Van Vleck 1928a}.
\end{quotation}
Van Vleck was thus ready enough to give up orbits once Heisenberg had shown the way. He failed to take this step on his own. 

The study of Van Vleck's paper illuminates  various aspects of the transition from the old quantum theory to matrix mechanics that tend to get obscured when one approaches these developments through, say, \citep{Kramers and Heisenberg 1925}. Most importantly perhaps, following \citep{Van Vleck 1924b, Van Vleck 1924c} rather than \citep{Kramers and Heisenberg 1925} or \citep{Born 1924}, we were able to give a transparent and explicit version of the derivation needed to show that the crucial Kramers dispersion formula reduces to the classical formula in the limit of high quantum numbers (see secs.\ 5.1--5.2 for the special case of a simple harmonic oscillator, sec.\ 6.2 for the generalization to arbitrary non-degenerate multiply-periodic systems, and sec.\ 7.1 for a closely analogous derivation of the Kramers formula in modern quantum mechanics). That Van Vleck confirmed the Kramers dispersion formula without relying on the Bohr-Kramers-Slater (BKS) theory makes it particularly clear that matrix mechanics grew directly out of dispersion theory and that BKS was mainly a sideshow (see sec.\ 4). The only element of the BKS theory used by Van Vleck is the concept of virtual oscillators. We saw that this concept actually predates BKS. `Virtual oscillators' was Bohr's new name for the substitute oscillators introduced into dispersion theory the year before and at Bohr's suggestion by \citet{Ladenburg and Reiche 1923}. In addition to popularizing the notion of virtual oscillators, BKS may have contributed to instilling skepticism about the electron orbits of the Bohr-Sommerfeld theory. In that sense, it might have helped Van Vleck had he embraced BKS more wholeheartedly. Overall, however, we argued that BKS  played no role in the breakthrough to matrix mechanics. The same is true for the broad acceptance of Einstein's light-quantum concept following the discovery of the Compton effect. Physicists working in dispersion theory, while accepting the Compton effect as decisive evidence for light quanta, happily continued to treat light as a wave phenomenon. 

What was it about dispersion theory that made it so important for the transition from the Bohr-Sommerfeld theory to the theory of Heisenberg's {\it Umdeutung} paper? As we suggested in the introduction of sec.\ 3, the answer is that the discrepancy between orbital frequencies and radiation frequencies---one of the most radical, if not {\it the} most radical aspect of the Bohr model of the atom---manifested itself glaringly and unavoidably in dispersion theory. The natural approach to adapting the successful classical dispersion theory of Lorentz and Drude to Bohr's new theory inevitably led to a dispersion formula with resonance poles at the orbital frequencies \citep{Sommerfeld 1915b, Debye 1915, Davisson 1916, Epstein 1922c}, whereas experiment clearly indicated that the resonance poles should be at the radiation frequencies, associated in Bohr's theory with transitions between orbits. Employing Einstein's quantum theory of radiation and Bohr's correspondence principle (in conjunction with techniques from celestial mechanics customized to the problems at hand) and building on pioneering work by \citet{Ladenburg 1921} and \citet{Ladenburg and Reiche 1923}, \citet{Kramers 1924a, Kramers 1924b} constructed a quantum formula for dispersion with resonance poles at the transition frequencies rather than at the orbital frequencies and claimed that this formula merged with the classical formula in the limit of high quantum numbers. \citet{Van Vleck 1924b, Van Vleck 1924c} was the first to publish an explicit proof that the Kramers quantum formula does indeed merge with the classical formula for dispersion in a general non-degenerate multiply-periodic system in the correspondence limit. The three key moves in translating the classical formula into a quantum-theoretical one were to (1) replace orbital frequencies by transition frequencies; (2) relate amplitudes to Einstein's $A$ coefficients; and (3) replace derivatives with respect to the action variable by difference quotients. The first move goes back to the embryonic version of the correspondence principle in \citep{Bohr 1913} \citep[pp.\ 274--275]{Heilbron and Kuhn}. \citet{Ladenburg 1921} introduced the second move. It was made more precise by Kramers and Van Vleck (cf.\ Jordan's remarks quoted in sec.\ 2.4). \citet{Born 1924} is usually credited with the third move and the rule for replacing derivatives by difference quotients is sometimes even called  ``Born's correspondence rule" \citep[p.\ 193]{Jammer 1966} or ``Born's discretizing rule" \citep[p.\ 181]{Cassidy}. It was found earlier, however, by both Kramers and Van Vleck (see the discussion at the end of sec.\  5.2).

The Kramers dispersion formula no longer contains any reference to the orbits of the Bohr-Sommerfeld theory, but only to transitions between them. This signaled to Heisenberg that orbits could be dispensed with altogether. Dispersion theory further told Heisenberg how to generate quantum formulae from classical formulae in his {\it Umdeutung} scheme. The procedure consisted of the same three moves listed above: one had to replace (1) classical frequencies (more specifically: the Fourier overtones of the classical mechanical motion) by quantum transition frequencies; (2) classical amplitudes associated with definite orbits by quantum transition amplitudes associated with pairs of stationary states;  and (3) derivatives by difference quotients. Dispersion theory also furnished the fundamental quantization condition for Heisenberg's new theory. Heisenberg formulated this condition by applying his {\it Umdeutung} procedure to the Bohr-Sommerfeld quantum condition, which was no longer acceptable because of its explicit reference to orbits. That Heisenberg's new condition also emerged in the high-frequency limit of the Kramers dispersion formula (see sec.\ 7.1) convinced him that he had found a sensible replacement for the Bohr-Sommerfeld condition. The relevant formula had been found in quantum dispersion theory before and was known as the Thomas-Kuhn(-Reiche) sum rule \citep{Thomas 1925, Kuhn 1925, Reiche and Thomas 1925}.  Van Vleck actually was the first to find this rule, even though he did not emphasize the result because he thought it was problematic (see sec.\ 3.5). According to Roger Stuewer (private communication), Van Vleck was nonetheless very proud of this achievement and used to mention it with pride to various colleagues in his later years. The Kramers dispersion formula and its corollary, the Thomas-Kuhn sum rule, are the critical physical ingredients in the first two sections of \citep{Heisenberg 1925c}, in which the {\it Umdeutung} procedure is motivated. Van Vleck was fully cognizant of these same ingredients by mid-1924. Van Vleck can thus truly be said to have been on the verge of {\it Umdeutung} in Minnesota in the summer of 1924.

\end{document}